\documentclass[letterpaper, 10pt]{article}
\usepackage[utf8]{inputenc}
\usepackage[top=1in, bottom=1in, left=1in, right=1in]{geometry}
\usepackage{float}
\usepackage{color}
\usepackage{tikz}
\usepackage{booktabs}
\usepackage{comment}

\usepackage{titling}
\RequirePackage[explicit]{titlesec}
\titlespacing*{\section}{0pt}{0.5em}{0.3pt}
\titlespacing*{\subsection}{0pt}{0.35em}{0pt}
\titlespacing*{\subsubsection}{0pt}{0.25em}{0pt}

\usepackage{caption}
\RequirePackage{fancyhdr}
\usepackage{lastpage}
\usepackage{empheq}

\usepackage{mathpazo}
\usepackage{xcolor}
\definecolor{GT}{RGB}{0, 0, 0}
\definecolor{Conv}{RGB}{247,37,133}
\definecolor{MC}{RGB}{114,9,183}
\definecolor{MVN}{RGB}{67,97,238}
\definecolor{FE}{RGB}{76,201,240}

\usepackage[
  colorlinks=true,
  linkcolor=blue,
  anchorcolor=blue,
  citecolor=blue,
  filecolor=blue,
  menucolor=blue,
  runcolor=blue,
  urlcolor=blue]{hyperref}
\usepackage{amsmath, amsthm, amssymb, amsfonts}
\usepackage{mathrsfs}
\usepackage{upgreek}
\usepackage{dsfont}

\usepackage{enumerate}
\usepackage{enumitem}

\usepackage{graphicx}
\usepackage{subcaption}

\usepackage{multirow}
\usepackage{multicol}

\usepackage[numbers, sort&compress]{natbib}

\makeatletter
\pretocmd\@maketitle
  {\def\@makefnmark{\hbox{\@textsuperscript{\normalfont\@thefnmark}}}}
  {}{\FAIL}
\makeatother
\makeatletter
\renewcommand*{\@fnsymbol}[1]{\ensuremath{\ifcase#1\or *\or \dagger\or \ddagger\or
   \mathsection\or \mathparagraph\or \|\or **\or \dagger\dagger
   \or \ddagger\ddagger \else\@ctrerr\fi}}
\makeatother

\title{Neural Optical Flow for Planar and Stereo PIV}

\author{
  Andrew I. Masker\thanks{Authors made an equal contribution.}, Ke Zhou$^*$, Joseph P. Molnar, and Samuel J. Grauer\thanks{Corresponding author: \href{mailto:sgrauer@psu.edu}{sgrauer@psu.edu}}\vspace*{.15em}\\
  {\small Department of Mechanical Engineering, Pennsylvania State University}}
\date{}

\begin{document}

\maketitle
\setcounter{footnote}{0}
\vspace*{-2em}

\begin{abstract}
\noindent 
Neural optical flow (NOF) offers improved accuracy and robustness over existing OF methods for particle image velocimetry (PIV). Unlike other OF techniques, which rely on discrete displacement fields, NOF parameterizes the physical velocity field using a continuous neural-implicit representation. This formulation enables efficient data assimilation and ensures consistent regularization across views for stereo PIV. The neural-implicit architecture provides significant data compression and supports a space--time formulation, facilitating the analysis of both steady and unsteady flows. NOF incorporates a differentiable, nonlinear image-warping operator that relates particle motion to intensity changes between frames. Discrepancies between the advected intensity field and observed images form the data loss, while soft constraints, such as integrated Navier--Stokes residuals, enhance accuracy and enable direct pressure inference from PIV images. Additionally, mass continuity can be imposed as a hard constraint for both 2D and 3D flows. Results from synthetic planar and stereo PIV datasets, as well as experimental planar data, demonstrate that NOF outperforms state-of-the-art wavelet-based OF, cross-correlation, and selected supervised machine learning methods. Beyond PIV, NOF could be used in conjunction with techniques like background-oriented schlieren, molecular tagging velocimetry, and other advanced measurement systems.\par\vspace{.5em}

\end{abstract}
\vspace*{2em}

\section{Introduction}
\label{sec:intro}
Enhanced processing techniques for particle image velocimetry (PIV) have significantly improved the accuracy and resolution of velocity measurements \cite{Westerweel2013}. PIV is a widely used, non-intrusive technique that captures 2D--3D velocity fields from images of a particle-laden flow exposed to a pulsed laser sheet or slab \cite{Raffel2018}. Tracer particles are advected between frames, illuminated by the laser, and imaged; sequential images are processed using a computer vision algorithm, cross-correlation (CC) algorithm, or similar to determine the particle displacement field. This field is converted to a velocity field, assuming a constant velocity between frames. Spatially resolved velocity fields obtained through PIV can reveal key flow features such as vortices, separation bubbles, and organized packets of coherent structures \cite{Scarano2012}. Furthermore, velocity may be used to calculate other fields like pressure, vorticity, strain rate, and Reynolds stresses, which are essential for building understanding, characterizing the performance of engineering devices, and validating computational models \cite{Adrian2005, Westerweel1997}. Despite the success of current methods, however, the need for more accurate measurements and greater resolution continues to drive the development of improved PIV techniques.\par

Cross-correlation is the standard method for extracting velocity fields from pairs of particle images in PIV. In this approach, images are divided into smaller regions called interrogation windows, and the correlation between windows in successive frames is computed to determine the average displacement of particles inside the window. The most probable displacement is identified by the peak correlation between windows across the image pair \cite{Westerweel2013}. A single velocity vector is generated for each interrogation window, and these windows are typically overlapped by 50--75\% to produce a denser vector field. Multi-resolution techniques are also employed to improve the performance of CC by estimating a coarse displacement field using large windows, warping one of the images accordingly, estimating a finer displacement field with smaller windows, and iterating this process \cite{Scharnowski2020, Huang1993, Theunissen2006}. Hu et al. \cite{Hu1998} found that optimal results are achieved when the smallest interrogation window is large enough to contain multiple particles but small enough to minimize excessive smoothing, typically ranging in size from $16 \times 16$ to $32 \times 32$ pixels. Nevertheless, even under optimal conditions, CC displacement yields fields with a vector spacing of about five pixels, causing velocity and gradient fields to appear smeared or low-pass filtered \cite{Hu1998, Kahler2012, Schmidt2019, Scarano2003}. Despite significant efforts to improve the resolution of CC-based measurements \cite{Hart1999, Takehara2000, Scarano2001, Susset2006, Kahler2012, Schanz2016}, modern implementations still produce relatively coarse deflection fields. Modern data-driven or machine-learning enhancements can lessen these issues at the cost of reduced generalizability and increased computational cost. Nevertheless, CC remains the most widely used technique for PIV analysis due to its computational efficiency, simplicity, and reliable performance.\par

Optical flow (OF) offers an alternative approach to PIV by computing a displacement field that warps an image taken at time $t$ to match the image from time $t + \Delta t$ \cite{Jassal2025}. OF algorithms can generate a dense velocity field, with one vector per pixel,\footnote{Although the number and arrangement of displacement vectors can vary in OF, it is standard to estimate one vector per pixel.} making them particularly effective for capturing flows with broadband spectral content and sharp gradients. However, classical OF methods are limited to handling small displacements \cite{Liu2008}, usually requiring a multi-resolution scheme to accommodate larger displacements \cite{Heitz2009, Champagnat2011}. These schemes can struggle with flows featuring both high gradients and a wide range of scales, such as turbulent shear layers or boundary layers \cite{Kahler2016}. Hybrid approaches that integrate OF with CC have been proposed to address these issues, but they introduce additional parameters to tune and increase computational complexity \cite{Alvarez2009, Heitz2009}. Since OF is inherently ill-posed, regularization is required to obtain a stable solution; in most cases, the regularization parameters are heuristically tuned for each flow scenario \cite{Corpetti2006}. Despite its advantages in accuracy and resolution, the adoption of OF for PIV has been limited due to its higher computational cost compared with CC and its sensitivity to spatio-temporal variations in laser output and out-of-plane motion, both of which amplify errors.\par

Wavelet-based OF (WOF) is the state of the art in OF for PIV. Introduced by Wu et al. \cite{Wu2000} and further developed by D{\'e}rian et al. \cite{Derian2013}, WOF represents the displacement field in the wavelet domain, taking advantage of the inherent multi-scale nature of the wavelet transform \cite{Mallat1999} to resolve turbulent flows with significant scale separation. Displacement estimates are refined using a coarse-to-fine approach, progressively optimizing wavelet coefficients across successive levels of the transform. Regularization is applied either by truncating the transform or adding a penalty term that promotes smoothness and stability. The wavelet coefficients are optimized via gradient descent, with spatial gradients in the penalty term computed directly in the wavelet domain \cite{Derian2013, KadriHarouna2013}. Schmidt and Sutton \cite{Schmidt2019, Schmidt2020} benchmarked WOF against state-of-the-art CC methods across various flow scenarios and particles-per-pixel (ppp) values, demonstrating that WOF consistently outperforms CC, especially at high wavenumbers and in capturing gradients, leading to more accurate vorticity fields. More recently, Jassal and Schmidt \cite{Jassal2023} and Nicolas et al. \cite{Nicolas2023} highlighted WOF's advantages in recovering near-wall velocity content.\par

The rapid adoption of machine learning (ML) in fluid dynamics has been driven by its success in other fields and the suitability of ML methods for image processing. These strengths make ML a natural fit for analyzing data from fluid diagnostics. One of the earliest breakthroughs in ML for fluid applications involved convolutional neural networks (CNNs), which excel at processing image data embedded in a low-dimensional, nonlinear manifold \cite{Krizhevsky2012, Goodfellow2014, Silver2016, Redmon2017, Lagemann2019}. Building on this foundation, CNNs have been applied to PIV processing. Rabault et al. \cite{Rabault2017} and Lee et al. \cite{Lee2017} developed PIV-DCNN, a CNN model for end-to-end motion estimation from particle images, mimicking CC in producing a sparse velocity field with one vector per interrogation window, e.g., $32 \times 32$ pixels in size. Cai et al. \cite{Cai2019a, Cai2019b} with their PIV-LiteFlowNet-en and other models augmented this approach to achieve dense motion estimation, generating one vector per pixel with accuracy comparable to Horn--Schunck OF \cite{Cai2019b}. However, supervised CNN-based algorithms require extensive pre-training on large, labeled datasets, typically comprising synthetic PIV images produced via direct numerical simulations (DNSs). This inevitably increases the computational cost of these approaches, due to dataset generation and the extensive training requirements. Additionally, the dependence on specific training data introduces the risk of overfitting and limits the generalizability of the trained models to unseen flow scenarios. As a result of these concerns, some unsupervised methods have been developed to reconstruct velocity from particle images. Zhang et al. \cite{Zhang2020} trained a CNN to output a velocity field; this field was used to warp input image pairs to match one another. This technique is conceptually similar to our method and to other unsupervised OF techniques \cite{Cai2021a, Qi2024}, with a primary distinction being the usage of a discrete CNN basis.\par

Neural-implicit fields are an emerging paradigm in computer vision and scientific ML that operate in an unsupervised manner \cite{Oechsle2021, Pumarola2021, Mildenhall2021, Tancik2022, Miotto2025}. Unlike many traditional methods that represent fields with a compact grid or set of overlapping basis functions, these networks, termed coordinate neural networks, represent variables as continuous functions of the input coordinates through one or more neural networks, providing significant data compression, particularly for spatio-temporally resolved fields \cite{Pumarola2021}. Exact partial derivatives of the networks can be efficiently computed via automatic differentiation, making these representations ideal for gradient-based optimization and data assimilation \cite{Raissi2019, Raissi2020, Miotto2025}, where the fields are optimized to satisfy governing differential equations. Unlike CNNs and other supervised models, coordinate neural networks can circumvent the need for labeled datasets, overcoming a key limitation of most ML methods used for PIV processing. Their versatility has led to applications in neural radiance fields for view synthesis \cite{Mildenhall2021, Tancik2022}, tomographic imaging \cite{Zang2021, Ruckert2022, Kelly2024, Molnar2025}, and data assimilation \cite{He2020, Patel2024, Liu2024}.\par

Recently, Qi et al. \cite{Qi2024} introduced a neural-implicit OF algorithm for PIV, demonstrating its ability to handle large displacements and enforce physical constraints through a Helmholtz decomposition. However, their method relies on ``HDNet,'' a differentiable Poisson solver trained on extensive synthetic datasets. This component reintroduces challenges typical of supervised learning and may limit generalizability, as seen in CNN-based PIV methods.\par

We present \emph{neural optical flow} (NOF), a framework that builds on the neural-implicit approach to modeling flow fields. NOF also incorporates a differentiable, nonlinear image-warping operator to estimate large displacements without the need for a multi-resolution scheme. The velocity field in world space is represented as a continuous function of space and time, making the method naturally extensible to higher-dimensional techniques like tomographic PIV. This space--time formulation offers substantial compression, reduces computational costs, and ensures seamless handling of steady and unsteady flows.\par

Our framework builds on the neural-implicit OF method of Qi et al. while eliminating the need for HDNet to enforce mass continuity. Where applicable, we impose continuity as a hard constraint through scalar or vector velocity potential fields; we can also include the Navier--Stokes equations into the optimization process, transforming NOF into a physics-informed neural network (PINN) \cite{Raissi2019}. Regularization is achieved by minimizing equation residuals, improving flow reconstruction from sparse measurements and enabling direct pressure inference from PIV images \cite{Cai2021a}. In addition to supporting data assimilation, NOF natively integrates stereo PIV by employing forward and inverse camera transforms to map between sensor and world coordinates. This allows us to use a single estimate of the velocity field across both perspectives, ensuring consistent regularization. NOF improves upon previous PINN-based methods, such as that of Cai et al. \cite{Cai2021a}, by moving beyond the linear OF equation, making our approach more effective for turbulent flows with a broad range of spectral content.\par

By combining the frameworks of Qi et al. \cite{Qi2024} and Cai et al. \cite{Cai2021a}, with our own innovations, we introduce a more accurate and robust OF technique. NOF is capable of performing PIV analyses within 3--25 minutes depending on the flow complexity, making the method usable in real world scenarios and competitive with other analysis tools, considering the improvements in accuracy an increased capabilities. We validate NOF on both synthetic and experimental PIV datasets, comparing its performance to conventional CC, a state-of-the-art wavelet-based OF method, and selected ML PIV algorithms. Our results demonstrate NOF's superior accuracy, robustness, and computational efficiency across a range of scenarios. The remainder of this paper is organized as follows. Section~\ref{sec:method} introduces the concepts underlying NOF. In Sec.~\ref{sec:cases}, we describe the synthetic and experimental flow cases used to evaluate PIV techniques. Numerical and experimental results are presented in Secs.~\ref{sec:results} and \ref{sec:exp results}, respectively, followed by our conclusions.\par

\section{Methodology}
\label{sec:method}
Neural optical flow estimates the fluid motion that warps the first image in a pair to match the second. To do this, it relies on coordinate transformations that relate physical motion in ``world space'' to apparent motion in the image, i.e., ``sensor space.'' The light captured by the cameras originates from a laser-illuminated 2D plane, $\mathcal{P}$, defined by a set of 3D points, $\mathbf{x} = (x_1, x_2, x_3) \in \mathcal{P}$, in world coordinates. Images are resolved in the sensor plane, $\mathcal{S}$, where $\mathbf{s} = (s_1, s_2) \in \mathcal{S}$ is a position in pixel units. A unique bijective mapping exists between $\mathcal{P}$ and $\mathcal{S}$. The forward mapping, commonly referred to as the camera transform, is defined as
\begin{equation}
    \label{equ:forward transform notation}
    \mathsf{\Psi} : \mathcal{P} \rightarrow \mathcal{S},
\end{equation}
and the inverse mapping, or inverse camera transform, is
\begin{equation}
    \label{equ:inverse transform notation}
    \mathsf{\Psi}^{-1} : \mathcal{S} \rightarrow \mathcal{P}.
\end{equation}
Examples of $\mathsf{\Psi}$ and $\mathsf{\Psi}^{-1}$ relevant to PIV are provided in \ref{app:camera}. Two additional quantities are essential: (1)~the magnification of the optical system for light from a point $\mathbf{x}$,
\begin{equation}
    \label{equ:magnification}
    M = \frac{f}{\left\lVert \mathbf{x} - \mathbf{c} \right\rVert_2 - f} \quad,
\end{equation}
where $\mathbf{c}$ denotes the camera's position, $f$ is the focal length of the lens, and $\lVert\cdot\rVert_2$ is the Euclidean norm, and (2)~the pixel pitch, $\psi$, defined as the reciprocal of the pixel size. With these elements in hand, we can describe the estimation of velocity fields from PIV data.\par

In the following sections, we introduce a differentiable image-warping operator and demonstrate its use in data loss functions for planar and stereo PIV. We then examine various regularization techniques, distinguishing between physics-\emph{inspired} and physics-\emph{informed} methods, and we explore different formulations of NOF.\par

\subsection{Neural Optical Flow}
\label{sec:method:NOF}
Optical flow is a computer vision method used to estimate the \emph{apparent} motion in a sequence of images based on changes in intensity. Over the years, many variations of OF have been developed, leveraging pixel-based methods, wavelets, or neural networks, as reviewed by Mendes et al. \cite{Mendes2022}. Our NOF technique builds on these approaches by employing a neural-implicit representation of the physical velocity field. Like other OF algorithms, NOF assumes that the brightness of a displaced feature is conserved between consecutive frames:
\begin{equation}
    \label{equ:NOF:conservation}
    I\mathopen{}\left(\mathbf{s},t\right) = I\mathopen{}\left(\mathbf{s} + \Delta \mathbf{s},t+ \Delta t\right).
\end{equation}
Here, $I$ is the image intensity at a sensor position $\mathbf{s}$ and time $t$; $\Delta\mathbf{s}$ and $\Delta t$ are the image displacement and time interval between frames, respectively.\par

In PIV, the displacement of a visible feature (usually a bright spot) in the image corresponds to the advection of a particle by the fluid in world space,
\begin{subequations}
    \label{equ:NOF:displacement transform}
    \begin{align}
        \Delta \mathbf{x} &= \int_{t}^{t+\Delta{}t} \underbrace{\frac{\mathrm{d}\widetilde{\mathbf{x}}\mathopen{} \left(\tau\right)}{\mathrm{d}\tau}}_{\mathbf{v}\mathopen{}\left(\widetilde{\mathbf{x}}, \tau\right)} \mathrm{d}\tau, \label{equ:NOF:displacement transform:advection}
        \intertext{where $\widetilde{\mathbf{x}}$ is the streamline starting from position $\mathbf{x}$ at time $t$ and $\mathbf{v}= (v_1, v_2, v_3)$ is the fluid velocity. The physical displacement is then projected onto the camera sensor,}
        \Delta \mathbf{s} &\equiv \underbrace{\mathsf{\Psi}\mathopen{}\left(\mathbf{x} + \Delta \mathbf{x}\right)}_{\mathbf{s} + \Delta \mathbf{s}} - \underbrace{\mathsf{\Psi}\mathopen{}\left(\mathbf{x}\right)}_{\mathbf{s}}. \label{equ:NOF:displacement transform:projection}
    \end{align}
\end{subequations}
OF aims to solve for the apparent displacement, $\Delta \mathbf{s}$, which evolves the image intensity $I$ from time $t$ to $t + \Delta t$. Once $\Delta \mathbf{s}$ is determined, it is mapped back to the two in-plane displacement components, $(\Delta x_1, \Delta x_2)$.\footnote{This assumes the laser sheet aligns with the  $(x_1,x_2)$ coordinates, otherwise a coordinate transformation is required.} The fluid velocity is then approximated as $\mathbf{v} \approx \Delta \mathbf{x}/\Delta t$.\par

Solving for the $\Delta \mathbf{s}$ field through Eq.~\eqref{equ:NOF:conservation} is a challenging optimization problem. A common simplification is to approximate the right-hand side with a first-order Taylor series expansion, evaluated at $(\mathbf{s}, t)$,
\begin{subequations}
    \label{equ:NOF:Taylor exp}
    \begin{align}
        I \approx&\, I + \frac{\mathrm{d}I}{\mathrm{d}t} \,\Delta t \label{equ:NOF:Taylor exp:1}\\
        =&\, I + \left( \frac{\Delta\mathbf{s}}{\Delta t} \cdot \nabla_\mathbf{s} I + \frac{\partial I}{\partial t}\right) \Delta t, \label{equ:NOF:Taylor exp:2}
    \end{align}
\end{subequations}
where $\partial I/\partial t$ is approximated by $[I\left(\mathbf{x}, t+\Delta t\right)-I\left(\mathbf{x}, t\right)]/\Delta t$ and $\nabla_\mathbf{s}$ is the gradient operator in sensor coordinates. As $\Delta t \rightarrow 0$, we get the linear OF equation,
\begin{equation}
    \label{equ:NOF:Taylor linear}
    \frac{\partial I}{\partial t} + \mathbf{v}_\mathcal{S} \cdot \nabla_\mathbf{s} I = 0,
\end{equation}
where $\mathbf{v}_\mathcal{S}$ is the apparent velocity in the image, which can be scaled by $(\psi M)^{-1}$ to estimate $\mathbf{v}$. While this linearization works well for small displacements, it breaks down as $\Delta \mathbf{s}$ or $\Delta t$ becomes large \cite{Heitz2009, Alvarez2009, Molnar2024}. PIV algorithms must effectively handle large displacements, so we adopt the generalized OF model from Eq.~\eqref{equ:NOF:conservation} for this work.\par

Motion estimation by OF is formulated as an optimization problem with a data loss term, expressed either in integral form,
\begin{subequations}
    \begin{align}
    \label{equ:NOF:data loss:int}
    \mathscr{J}_\mathrm{data} &= \frac{1}{\left\lvert\mathcal{S}\right\rvert}\int_\mathcal{S} \left[I\mathopen{}\left(\mathbf{s},t\right) -I\mathopen{}\left(\mathbf{s} + \Delta\mathbf{s},t+\Delta{}t\right)\right]^2 \mathrm{d}\mathbf{s},
    \intertext{or differential form,}
    \label{equ:NOF:data loss:diff}
    \mathscr{J}_\mathrm{data} &= \frac{1}{\left\lvert\mathcal{S}\right\rvert}\int_\mathcal{S} \left(\frac{\partial I}{\partial t} + \frac{\Delta \mathbf{s}}{\Delta t} \cdot \nabla_\mathbf{s} I\right)^2\mathrm{d}\mathbf{s}.
    \end{align}
\end{subequations}
The latter assumes the linearization in Eq.~\eqref{equ:NOF:Taylor exp}. Either way, the objective is to find the displacement field, $\Delta \mathbf{s}$, that minimizes $\mathscr{J}_\mathrm{data}$. This is an ill-posed inverse problem with no unique solution, having only one piece of information from each pixel (i.e., the intensity) and two unknown parameters, $(\Delta s_1, \Delta s_2)$, which is known as the ``aperture problem'' \cite{Ullman1979}. To ensure a unique, physically plausible solution, most OF techniques introduce an explicit penalty term, $\mathscr{J}_\mathrm{penalty}$, which promotes desirable characteristics, such as spatial smoothness, in the solution. The total loss function is a weighted sum of the data and penalty terms,
\begin{equation}
\label{equ:NOF:optimization:subfinal}
    \mathscr{J}_\mathrm{total} = \mathscr{J}_\mathrm{data} + \lambda \mathscr{J}_\mathrm{penalty} 
\end{equation}
The weighting parameter $\lambda$ must be carefully selected to balance the loss terms. In this work, a near-optimal value is identified using L-curve analysis \cite{Hansen1992}. We find that NOF is relatively robust to $\lambda$ near optimality, with small changes to the method or dataset requiring little to no additional tuning. Future work will explore auto-weighting strategies \cite{Wang2021} and Bayesian optimization \cite{Calvetti2018}. The optimal displacement field, $\Delta\mathbf{s}^*$, is obtained by minimizing the total loss,
\begin{equation}
    \label{equ:NOF:optimization:final}
    \Delta\mathbf{s}^* = \mathrm{arg}\,\underset{\Delta\mathbf{s}}{\mathrm{min}} \left.\mathscr{J}_\mathrm{total}\right..
\end{equation}\par

Neural optical flow is distinguished by its use of a neural-implicit representation of the velocity field. Rather than relying on pixel- or wavelet-based methods, a neural network, $\mathsf{N}$, models the fluid velocity as a continuous function of space and time,
\begin{equation}
    \label{equ:NOF:flow}
    \mathsf{N} : \left(\mathbf{x}, t\right) \mapsto \mathbf{v}.
\end{equation}
In other words, $\mathsf{N}$ is a network that takes $\mathbf{x}$ and $t$ as inputs and outputs $\mathbf{v}$. Additional variables like pressure can also be outputted, as demonstrated in Sec.~\ref{sec:results:cylinder flow}. Details of the network architecture are provided in \ref{app:arch}. Given a velocity field estimated by $\mathsf{N}$, the displacement and corresponding objective loss are computed through the following steps:
\begin{itemize}
    \item A point $\mathbf{s} \in \mathcal{S}$ on the sensor is mapped to a position $\mathbf{x} \in \mathcal{P}$ in the laser plane using the inverse camera transform, $\mathsf{\Psi}^{-1}$. A displacement, $\Delta \mathbf{x}$, is then computed by numerically integrating the velocity field, $\mathbf{v}$, via Eq.~\eqref{equ:NOF:displacement transform:advection}.
    
    \item The displaced point, $\mathbf{x} + \Delta\mathbf{x}$, is projected onto the sensor plane to yield $\Delta \mathbf{s}$ using Eq.~\eqref{equ:NOF:displacement transform:projection}.
    
    \item The image intensities $I(\mathbf{s}, t)$ and $I(\mathbf{s} + \Delta \mathbf{s}, t + \Delta t)$ are sampled to evaluate the integral-form data loss in Eq.~\eqref{equ:NOF:data loss:int}.
\end{itemize}
A detailed explanation and visualization of each step is provided in the following section.\par

The NOF framework offers several advantages over existing OF methods. First, numerical integration of the velocity field allows it to handle large displacements without requiring a multi-resolution scheme \cite{Horn1981, Corpetti2002, Corpetti2006, Cai2019a, Cai2019b}. Second, NOF extends naturally to stereo PIV, as discussed in Sec.~\ref{sec:method:stereo}: the neural network represents all three velocity components in physical space and optimizes across both imaging perspectives simultaneously. Third, the method leverages a $C^\infty$ neural representation, enabling continuous regularization through automatic differentiation, as outlined in Sec.~\ref{sec:method:physics-inspired}. Implicit regularization is also achieved by tuning the spectral content embedded in the network's Fourier encoding, per \ref{app:arch}. Lastly, both weak- and hard-form physics-based constraints are introduced to refine velocity estimates and infer additional flow states, such as pressure, which is described in Sec.~\ref{sec:method:physics-based}.\par

\subsection{Data Loss for Planar PIV}
\label{sec:method:planar}
A key element of any OF technique is the warping operator, as the quality of the velocity field is assessed by warping the first image to approximate the second one. An ideal warping operator, $\mathsf{W}$, can be defined in terms of the sets
\begin{align}
    \label{equ:warp:set}
    \mathcal{I}(t) = \left\{I\mathopen{}\left(\mathbf{s}, t\right) \mid \mathbf{s} \in \mathcal{S}\right\} \quad\text{and}\quad
    \delta \mathcal{S} = \left\{\Delta \mathbf{s}\mathopen{}\left(\mathbf{s}\right) \mid \mathbf{s} \in \mathcal{S}\right\},
\end{align}
where $\mathcal{I}$ represents the imaged intensity field at time $t$ across the sensor domain, $\mathcal{S}$, and $\delta \mathcal{S}$ is the deflection field. The warping operator acts on this set,
\begin{equation}
    \label{equ:warp:ideal}
    \mathsf{W} : \left(\mathcal{I}, \delta \mathcal{S}\right) \mapsto \mathcal{I}^\prime,
\end{equation}
where $\mathcal{I}^\prime$ is the image warped by the displacement field $\delta \mathcal{S}$.\par

The linear OF warping operator, commonly used in OF algorithms, employs Eq.~\eqref{equ:NOF:Taylor linear} to describe changes to the initial image. Discrete derivatives of $I$, evaluated at the pixel centroids, are combined with the estimated velocity field to generate the warped image. This warped image is then compared to the experimental data, forming the objective loss in Eq.~\eqref{equ:NOF:data loss:diff}.\par

The nonlinear warping operators used in conjunction with Eq.~\eqref{equ:NOF:data loss:int} do not rely on such approximations. Nonlinear warping can be categorized into two classes. The first, more general approach advects the intensity at each position $\mathbf{s}$ to $\mathbf{s} + \Delta \mathbf{s}$, with the warped image $\mathcal{I}^\prime$ comprising the sum of deposited intensities at these new locations. However, this formulation is non-injective and not invertible, as illustrated by the scenario where two points in the first frame advect to the same location in the second frame.\par

Many optimization schemes call for an invertible mapping. This second class of warping assumes injectivity, ensuring that the operator satisfies
\begin{equation}
    \mathsf{W}\mathopen{}\left(\mathcal{I}, \delta \mathcal{S}\right) = \mathcal{I}^\prime \quad\text{and}\quad \mathcal{I} = \mathsf{W}\mathopen{}\left(\mathcal{I}^\prime, -\delta \mathcal{S}\right).
\end{equation}
With the added assumption of surjectivity, every point $\mathbf{s} \in \mathcal{S}$ in the first frame maps to a unique point in the second frame, allowing the integrals in Eq.~\eqref{equ:NOF:data loss:int} to be approximated to arbitrary precision. In this work, we adopt this assumption, implying that all positions in $\mathcal{S}$ are advected to distinct locations on the sensor in the second frame. Our method remains robust even when these assumptions are not strictly met.\par

\begin{figure*}[ht!]
    \centering\vspace*{-0.66em}
    \includegraphics[width=0.9\textwidth]{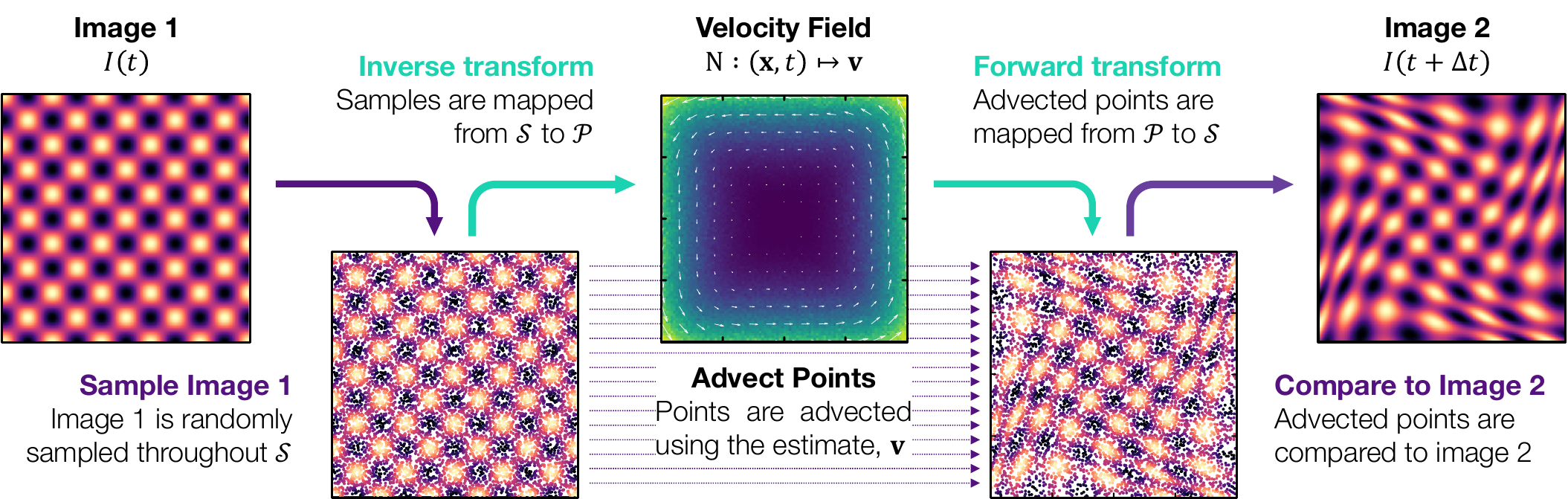}
    \vspace*{.5em}
    \caption{Schematic of the image warping procedure used in NOF. Points are sampled in $\mathcal{I}(t)$, advected to $t + \Delta t$ via integration of $\mathbf{v}$, and compared against the corresponding intensities in $\mathcal{I}(t + \Delta t)$.}
    \vspace*{-0.66em}
    \label{fig:warping}
\end{figure*}

Our warping scheme involves five steps, illustrated in Fig.~\ref{fig:warping}. First, a set of points is uniformly sampled from the sensor domain, $\mathcal{S}$. These points, $\{\mathbf{s}_1, \mathbf{s}_2, \dots\}$, are then mapped to world coordinates using the inverse camera transform, $\mathbf{x}_i = \mathsf{\Psi}^{-1}(\mathbf{s}_i)$, as detailed in \ref{app:camera}. Next, the points are advected from $\mathbf{x}_i$ to $\mathbf{x}_i + \Delta \mathbf{x}_i$ using a forward Euler scheme to approximate the integral in Eq.~\eqref{equ:NOF:displacement transform:advection}. Numerical integration of the continuous velocity field is crucial for managing large displacements; unless otherwise noted, we use a time step of $\Delta t/5$, which proved stable across all tested flow scenarios. Testing with a step size of $\Delta t$ showed inferior accuracy compared to our multi-step approach. While this study uses a forward Euler scheme, future work may explore higher-order integration methods.\par

In the fourth step, the final locations of the points are transformed back to sensor coordinates using the camera transform,
\begin{equation}
    \mathbf{s}_i + \Delta\mathbf{s}_i = \mathsf{\Psi}\mathopen{}\left(\mathbf{x}_i + \Delta \mathbf{x}_i\right).
\end{equation}
Next, pixel intensities are sampled from the first image at $\mathbf{s}_i$ and from the second image at $\mathbf{s}_i + \Delta \mathbf{s}_i$. When the displacement field is accurate, these intensity values should match. Intensities are retrieved via interpolation of the discrete image data: $\mathbf{I} = \{I_{i,j} \mid i = 1, \dots m, j = 1, \dots n\}$, where $I_{i,j}$ is the intensity of the $(i,j)$th pixel in an $m\times n$ image. While both bilinear and bicubic interpolation schemes were tested, bilinear interpolation is used throughout this work due to its lower computational cost and comparable accuracy across all cases. The interpolation operator is defined as
\begin{equation}
    \mathsf{B} : \left[\mathbf{I}\mathopen{}\left(t\right), \mathbf{s}\right] \mapsto \hat{I}\mathopen{}\left(\mathbf{s}, t\right),
\end{equation}
where $\hat{I}$ is the interpolated estimate of the continuous intensity field, $I$. Further details about the interpolation scheme are provided in \ref{app:interp}.\par

Lastly, in the fifth step, we use Monte Carlo sampling to evaluate an integral-form data loss,
\begin{subequations}
    \begin{gather*}
        \label{equ:planar NOF data loss}
        \mathscr{J}_\mathrm{data}^{k} = \frac{1}{\left\lvert\mathcal{S}_k\right\rvert}\int_{\mathcal{S}_k}\mathopen{}\left\{\mathsf{B}\mathopen{}\left[\mathbf{I}\mathopen{}\left(t\right), \mathbf{s}\right] - \mathsf{B}\mathopen{}\left[\mathbf{I}\mathopen{}\left(t + \Delta t\right), \mathbf{s} + \Delta \mathbf{s}_k \right]\right\}^2 \mathrm{d}\mathbf{s},
        \intertext{where $k$ denotes the camera index (to allow for multi-camera setups),}
        \Delta \mathbf{s}_k = \mathsf{\Psi}_k\mathopen{}\left[ \mathsf{\Psi}_k^{-1}\mathopen{}\left({\mathbf{s}}\right) + \Delta \mathbf{x}_k\right] - \mathbf{s},
    \end{gather*}
\end{subequations}
and $\Delta \mathbf{x}_k$ is the solution to Eq.~\eqref{equ:NOF:displacement transform:advection} initialized at $\mathsf{\Psi}_k^{-1}(\mathbf{s})$. For planar imaging, only a single camera is used, and the camera index can be omitted from this notation. However, capturing out-of-plane velocity components requires multiple cameras or auxiliary devices, such as endoscopes or fibers. In such cases, a separate data loss must be computed for each image, as discussed in the next section.\par

\subsection{Data Loss for Stereo PIV}
\label{sec:method:stereo}
Three-dimensional effects are common in turbulent flows, vortex-dominated flows, unstable shear flows, and more. Stereoscopic PIV (stereo PIV) uses two cameras to simultaneously capture particles within the measurement plane from distinct perspectives. Data from these views allow for the extraction of out-of-plane fluid motion and helps to reduce perspective errors for in-plane motion \cite{Prasad1993}. This two-dimensional, three-component (2D3C) velocimetry technique has become essential for accurately measuring velocity in flows with significant vortical content or other out-of-plane motions.\par

\begin{figure}[ht]
    \centering\vspace*{-0.66em}
    \includegraphics[width=\linewidth]{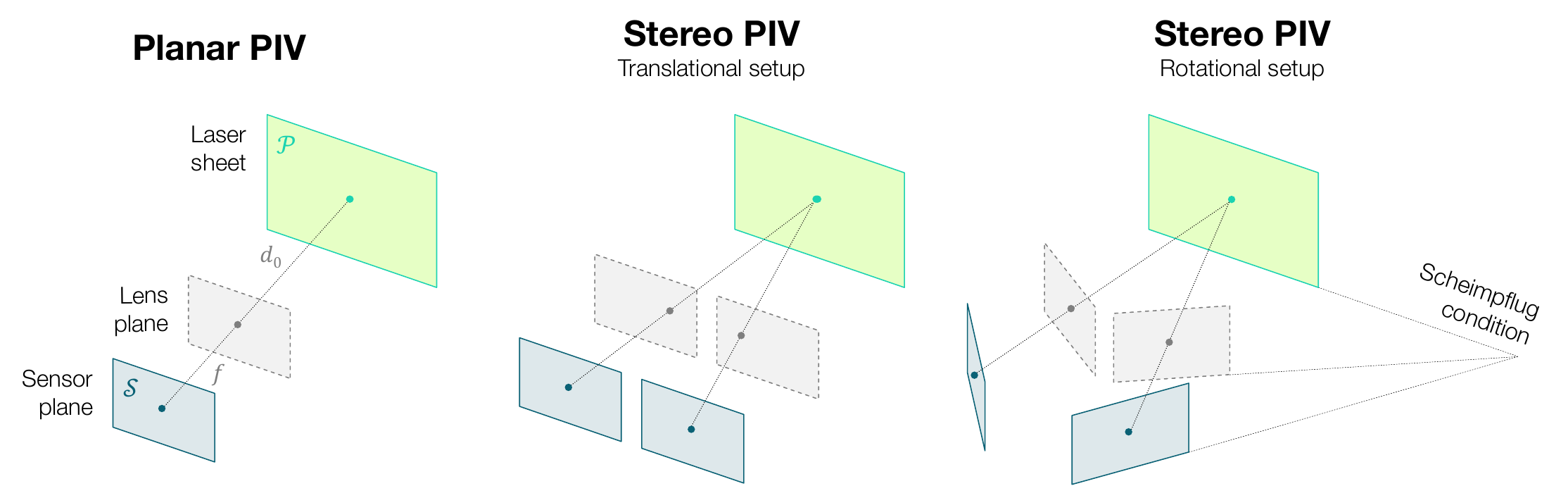}
    \vspace*{.33em}
    \caption{PIV setups: (left) planar, (middle) translational stereo, and (right) rotational stereo. Green planes represent the laser sheet, cobalt planes indicate the sensors, and gray planes denote the lenses. Diagrams not to scale.}
    \vspace*{-0.66em}
    \label{fig:PIV}
\end{figure}

Most stereo PIV setups follow a translational or rotational configuration \cite{Prasad2000}. Figure~\ref{fig:PIV} shows a graphical overview of planar and stereo PIV. In the translational version, both cameras have parallel optical axes that are perpendicular to the laser sheet. This setup is easy to implement, provides uniform magnification across the region of interest, and ensures a sharp focus, similar to single-camera planar PIV. However, the cameras in this configuration have a limited ``common area'' (i.e., the region of the laser sheet visible to both cameras). Expanding this area requires either (1) moving the cameras closer together, sacrificing some accuracy in the out-of-plane velocity component, or (2) translating the sensors relative to the lenses, which is challenging in many commercial cameras. Additionally, lens vignetting becomes more pronounced as the sensors move off-axis, especially at the large offsets needed for accurate out-of-plane motion estimation.\par

The rotational setup addresses these challenges by turning both cameras to point them towards a common point in the region of interest. Scheimpflug adapters are typically used to maintain a sharp focus from these oblique angles. This configuration overcomes the off-axis angle limitations of the translational method, reducing vignetting and enabling accurate out-of-plane motion estimation over a larger area. However, rotational setups introduce non-uniform magnification across the field of view, which complicates camera calibration and data processing. Both configurations are discussed in detail by Prasad \cite{Prasad2000}.\par

In this work, we demonstrate NOF using the translational configuration for simplicity. However, our algorithm is equally applicable to the rotational configuration with appropriate adjustments to the camera transform. Details of the translational and rotational transforms are provided in \ref{app:camera:stereo}.\par

Traditional stereo PIV processing computes a displacement field for each perspective using OF, CC, or a similar technique. These fields are then mapped onto the laser plane in world coordinates, enabling the estimation of both in-plane and out-of-plane motion, as described by Eq.~\eqref{equ:stereo:displacement transform} in \ref{app:camera:stereo}. In contrast, stereo NOF performs the inversion directly in world space. Our stereo data loss follows the same structure as Eq.~\eqref{equ:planar NOF data loss}, with points sampled on the sensors, transformed into world coordinates, advected, and then mapped back to image space. The key difference is that stereo NOF forms a stereo data loss, with one component per camera, using their respective forward and inverse camera transforms:
\begin{equation}
    \label{equ:stereo:NOF}
    \mathscr{J}_\mathrm{data} = \mathscr{J}_\mathrm{data}^\mathrm{L} + \mathscr{J}_\mathrm{data}^\mathrm{R}.
\end{equation}
Here, $\mathscr{J}_\mathrm{data}^\mathrm{L}$ and $\mathscr{J}_\mathrm{data}^\mathrm{R}$ are loss terms for the left and right cameras, respectively, both of which have dedicated $\mathsf{\Psi}$ and $\mathsf{\Psi}^{-1}$ operators.\par

Although our measurement is planar, advection by the 2D3C velocity field can move points out of the plane\slash laser sheet. In principle, sampling of the network, $\mathsf{N}$, could be performed throughout a thin volume surrounding the laser sheet, within which particles are illuminated by a Gaussian laser slab of finite thickness. However, in our tests, sampling within the central $x_1{-}x_2$ plane yielded the best results. Future work will explore improved sampling schemes to more accurately capture out-of-plane effects.\par

\subsection{Physics-Inspired Regularization}
\label{sec:method:physics-inspired}
An explicit penalty term is often included in the aggregate loss to close the OF equations and promote desirable solution characteristics. Most OF regularization methods use ``physics-inspired'' penalties that do not strictly align with the governing equations but encourage properties like smoothness or a known correlation length scale \cite{Kaipio2006}. NOF can apply the same penalty terms as other OF algorithms for PIV but does so in a \emph{truly} continuous form. For comparison, WOF relies on gradient operators tied to the discrete signal space associated with the wavelet transform. Moreover, we regularize the flow velocity within the measurement plane in world space, $\mathcal{P}$, rather than sensor space. This ensures consistent regularization for both views in stereo PIV and facilitates the more comprehensive, physics-based penalties discussed in the next section.\par

A common penalty choice is the first-order Tikhonov regularization method introduced to OF by Horn and Schunck \cite{Horn1981},
\begin{equation}
    \label{equ:reg:Tik}
    \mathscr{J}_\mathrm{penalty} = \frac{1}{\left|\mathcal{P}\right|} \int_\mathcal{P} \left\lVert \nabla_\mathbf{x} \mathbf{v} \right\rVert_2^2 \mathrm{d}\mathbf{x}.
\end{equation}
Tikhonov regularization promotes \emph{globally} smooth solutions in the $L^2$ sense. Another common choice is total variation (TV) regularization, which minimizes the overall gradient content to promote \emph{piecewise} smooth solutions that can sustain sharp discontinuities, i.e., enforcing smoothness in the $L^1$ sense,
\begin{equation}
    \label{equ:reg:TV}
    \mathscr{J}_\mathrm{penalty} = \frac{1}{\left|\mathcal{P}\right|} \int_\mathcal{P} \left\lVert \nabla_\mathbf{x} \mathbf{v} \right\rVert_1 \mathrm{d}\mathbf{x}.
\end{equation}
Note that $\lVert \cdot \rVert_1$ is the Manhattan norm, which is non-differentiable and typically requires a specialized optimization technique. Higher-order regularization terms, such as
\begin{equation}
    \label{equ:reg:third-order}
    \mathscr{J}_\mathrm{penalty} = \frac{1}{\left|\mathcal{P}\right|} \int_\mathcal{P} \left(\frac{\partial^3 v_1 }{\partial x_1^3}\right)^2 + \left(\frac{\partial^3 v_2 }{\partial x_2^3}\right)^2 \mathrm{d}\mathbf{x},
\end{equation}
can improve velocity estimation without excessively smoothing the field or penalizing divergence and curl \cite{Schmidt2019, KadriHarouna2013, Schmidt2020}. A particularly effective approach for 2D velocity fields is div--curl regularization,
\begin{equation}
    \label{equ:reg:div-curl}
    \mathscr{J}_\mathrm{penalty} = \frac{1}{\left|\mathcal{P}\right|} \int_\mathcal{P} \left\lVert\nabla_\mathbf{x} \left(\nabla_\mathbf{x} \cdot \mathbf{v}\right) \right\rVert_2^2 + \left\lVert \nabla_\mathbf{x} \left(\nabla_\mathbf{x} \times \mathbf{v}\right) \right\rVert_2^2 \mathrm{d}\mathbf{x},
\end{equation}
which aims to smooth the flow by penalizing gradients of divergence and curl while preserving the structure of vortical and compressive regions \cite{Corpetti2002, Corpetti2006, Yuan2007, Schmidt2020}.\par

Penalizing higher-order derivatives in an explicit penalty term often introduces numerical sensitivities that challenge optimization algorithms and increase computational costs. In NOF, the network architecture provides implicit regularization, as outlined in \ref{app:arch}. Adjusting the Fourier encoding influences the spectral content that the neural network can learn. The bandwidth of the network may be tailored to the flow, e.g., based on an estimated Reynolds number or known spectral properties \cite{Jin2024}. Systematic testing demonstrated that implicit regularization is both more accurate and efficient than explicit methods in NOF. Therefore, we employ implicit regularization throughout this work unless we state otherwise.\par

\subsection{Physics-Based Constraints}
\label{sec:method:physics-based}
Explicit physics-based constraints enhance the accuracy and physical consistency of solutions. While some constraints can be applied in other OF methods (typically linear constraints like mass continuity for incompressible flows \cite{Corpetti2006, Stark2013, Chong2022}), NOF offers distinct advantages. Specifically, our approach enables time-resolved reconstructions that include the Navier--Stokes equations. Additionally, NOF's space--time representation allows us to exploit sparsity in the time domain: an advantage not utilized by existing OF methods.\par

Many flows are incompressible, making it advantageous to represent the velocity field in a way that inherently satisfies the continuity equation,
\begin{equation}
    \label{equ:physics constraint:div-free}
    \frac{\partial \rho}{\partial t} + \nabla_\mathbf{x} \cdot \left(\rho \mathbf{v}\right) = 0 \Rightarrow \nabla_\mathbf{x} \cdot \mathbf{v} = 0.
\end{equation}
This is straightforward to implement in NOF. For example, a 2D divergence-free velocity field can be represented using a scalar potential, $\varphi$. By redefining the network output as $\mathsf{N} : (\mathbf{x}, t) \mapsto \varphi$, the velocity field $\mathbf{v}$ can be constructed as
\begin{equation}
    \label{equ:physics constraint:div-free:scalar}
    \mathbf{v} = \left( -\frac{\partial \varphi}{\partial y}, \, \frac{\partial \varphi}{\partial x}\right)
\end{equation}
using automatic differentiation. Taking the divergence of this field confirms compliance with the continuity equation,
\begin{equation}
    \nabla_\mathbf{x} \cdot \mathbf{v} = -\frac{\partial^2 \varphi}{\partial x \,\partial y} + \frac{\partial^2 \varphi}{\partial y \,\partial x} = 0.
\end{equation}
This reparametrization guarantees that all reconstructed velocity fields satisfy the divergence-free condition while maintaining a computational cost comparable to that of a soft constraint. Although this approach is limited to 2D flows, continuity can still be embedded as a hard constraint in 3D settings via a vector potential field,
\begin{equation}
    \boldsymbol\upvarphi = \left(\varphi_1 , \varphi_2 , \varphi_3\right).
\end{equation}
The network maps $(\mathbf{x}, t)$ to $\boldsymbol\upvarphi$ and the velocity field is given by
\begin{equation}
    \label{equ:physics constraint:div-free:vector}
    \mathbf{v} = \nabla_\mathbf{x} \times \boldsymbol\upvarphi,
\end{equation}
once again ensuring that the velocity field remains divergence-free,
\begin{equation}
    \nabla_\mathbf{x} \cdot \left(\nabla_\mathbf{x} \times \boldsymbol\upvarphi\right) = 0.
\end{equation}
In practice, two components of $\boldsymbol\upvarphi$ may be sufficient, with $\varphi_3 = 0$. Further research is needed to assess this approach for 3D3C measurements, such as tomographic PIV \cite{Scarano2012} and Lagrangian particle tracking \cite{Schanz2016}.

In addition to hard physics constraints, we incorporate soft constraints by including the residuals of the 2D Navier--Stokes equations to enable \emph{direct} time-resolved pressure estimation from PIV data for 2D flows:
\begin{equation}
    \label{equ:physics constraint:NS}
    \mathscr{J}_\mathrm{penalty} = \frac{1}{\left|\mathcal{P} \times \mathcal{T}\right|} \int_\mathcal{T} \int_\mathcal{P} \underbrace{\vphantom{\left(\frac{\partial}{\partial}\right)_1^1}\left(\nabla_\mathbf{x} \cdot \widehat{\mathbf{v}}\right)^2}_\text{continuity} + \underbrace{\left\lVert\frac{\partial \widehat{\mathbf{v}}}{\partial t} + \widehat{\mathbf{v}} \cdot \nabla_\mathbf{x} \widehat{\mathbf{v}}  + \nabla_\mathbf{x} \widehat{p} - \frac{1}{Re}\nabla^2\widehat{\mathbf{v}}\right\rVert_2^2}_\text{momentum conservation} \mathrm{d}\mathbf{x} \,\mathrm{d}t.
\end{equation}
In this expression, $\mathcal{T}$ is the time domain of the experiment while $\widehat{\mathbf{v}} = \mathbf{v}/v_0$ and $\widehat{p} = p/(\rho v_0^2)$ represent the dimensionless velocity and pressure, respectively, and $v_0$ and $\ell_0$ are the characteristic velocity and length scales. The reference Reynolds number is
\begin{equation}
    Re = \frac{v_0 \,\ell_0}{\nu},
\end{equation}
where $\nu$ is the kinematic viscosity. Note that the mass continuity residuals in Eq.~\eqref{equ:physics constraint:NS} are redundant when using one of the hard constraints presented above. Incorporating governing equation residuals into the loss effectively transforms NOF into a PINN, where the estimated velocity field not only aligns with PIV data but also approximately solves the Navier--Stokes equations. This enhances the accuracy of velocity estimates and enables simultaneous pressure inference. However, it only applies to 2D flows because planar cross-sections of fully-3D flows do not satisfy the 2D equations, and there is too much uncertainty for a 3D solution. In such cases, 3D measurements are required for effective data assimilation \cite{Zhou2023, Zhou2024}.\par

\subsection{Overview of NOF Variants}
\label{sec:method:NOF variants}
The preceding subsections outlined our NOF framework and its various configurations. NOF variants differ based on the imaging system (planar or stereo), the incorporation of temporal information (static or time-resolved reconstructions), and the use of constraints (hard physics-based constraints, soft regularization, or physics-informed loss terms). Each variant is tailored to specific flow conditions, offering distinct advantages for different experimental settings. Below, we summarize the key NOF variants and their respective use cases.
\begin{itemize}
    \item \texttt{NOF} (standard NOF): The closest analog to conventional OF, the vanilla formulation of NOF maps spatial coordinates to velocity, $\mathbf{x} \mapsto \mathbf{v}$. Soft regularization terms (Eqs.~\eqref{equ:reg:Tik}--\eqref{equ:reg:div-curl}) may be incorporated as needed. This formulation is broadly applicable to planar PIV datasets and can be used in stereo PIV by processing each view independently before recombination.

    \item \texttt{NOF-HD} (hard divergence): This variant enforces incompressibility as a hard constraint by mapping spatial coordinates to a scalar or vector potential, $\mathbf{x} \mapsto \varphi$ or $\boldsymbol\upvarphi$, from which a divergence-free velocity field is constructed via Eq.~\eqref{equ:physics constraint:div-free:scalar} or \eqref{equ:physics constraint:div-free:vector}, respectively. NOF-HD is well suited for incompressible, quasi-planar flows, including the planar PIV cases in Secs.~\ref{sec:results:2D HIT} and \ref{sec:results:cylinder flow}. Its application to fully 3D flows is more complex due to the lack of out-of-plane gradients in planar and stereo PIV; future work will explore its performance in 3D flow reconstructions.

    \item \texttt{NOF-TR} (time-resolved): This extension incorporates temporal information by mapping spatio-temporal coordinates, $(\mathbf{x}, t)$, to velocity. Training a single neural network on an entire time series improves temporal coherence and reduces computational cost compared to solving each frame independently. NOF-TR is evaluated in Sec.~\ref{sec:results:2D HIT} to assess its impact on pointwise errors and spectral content, but it is omitted from other sections where time-resolved results provide limited additional insights.

    \item \texttt{NOF-phys} (physics-based): Physics-informed NOF incorporates governing equations (e.g., Navier--Stokes and mass continuity) into the loss function, transforming NOF into a PINN. The network maps $(\mathbf{x}, t) \mapsto (\mathbf{v}, p)$, simultaneously recovering velocity and pressure fields. By embedding physical constraints, NOF-phys improves velocity estimation and enables direct pressure inference from PIV data. It is applied in Secs.~\ref{sec:results:2D HIT} and \ref{sec:results:cylinder flow} but is omitted from stereo tests (Sec.~\ref{sec:results:3D HIT}) due to the lack of out-of-plane gradients, corresponding to an underdetermined problem. NOF-phys is particularly well suited for planar and non-swirling axisymmetric flows and can be used for both steady and time-resolved datasets; however, we only consider the time-resolved formulation in this study.

    \item \texttt{Stereo NOF}: Any of the NOF formulations above can be extended to stereo PIV by jointly optimizing the velocity field using data losses for both camera views. This approach eliminates errors associated with independent processing and the stereo recombination step, leading to more consistent and physically accurate reconstructions.
\end{itemize}

While these NOF variants are presented as discrete categories for clarity, they are not mutually exclusive. Many combinations are possible, such as time-resolved physics-informed reconstructions with hard divergence constraints. This study focuses on the fundamental NOF variants outlined above, but future work will explore hybrid formulations to further enhance reconstruction accuracy and efficiency.\par

\section{Flow Scenarios and Metrics}
\label{sec:cases}
\begin{figure*}[ht!]
    \centering\vspace*{-0.66em}
    \includegraphics[width=.95\textwidth]{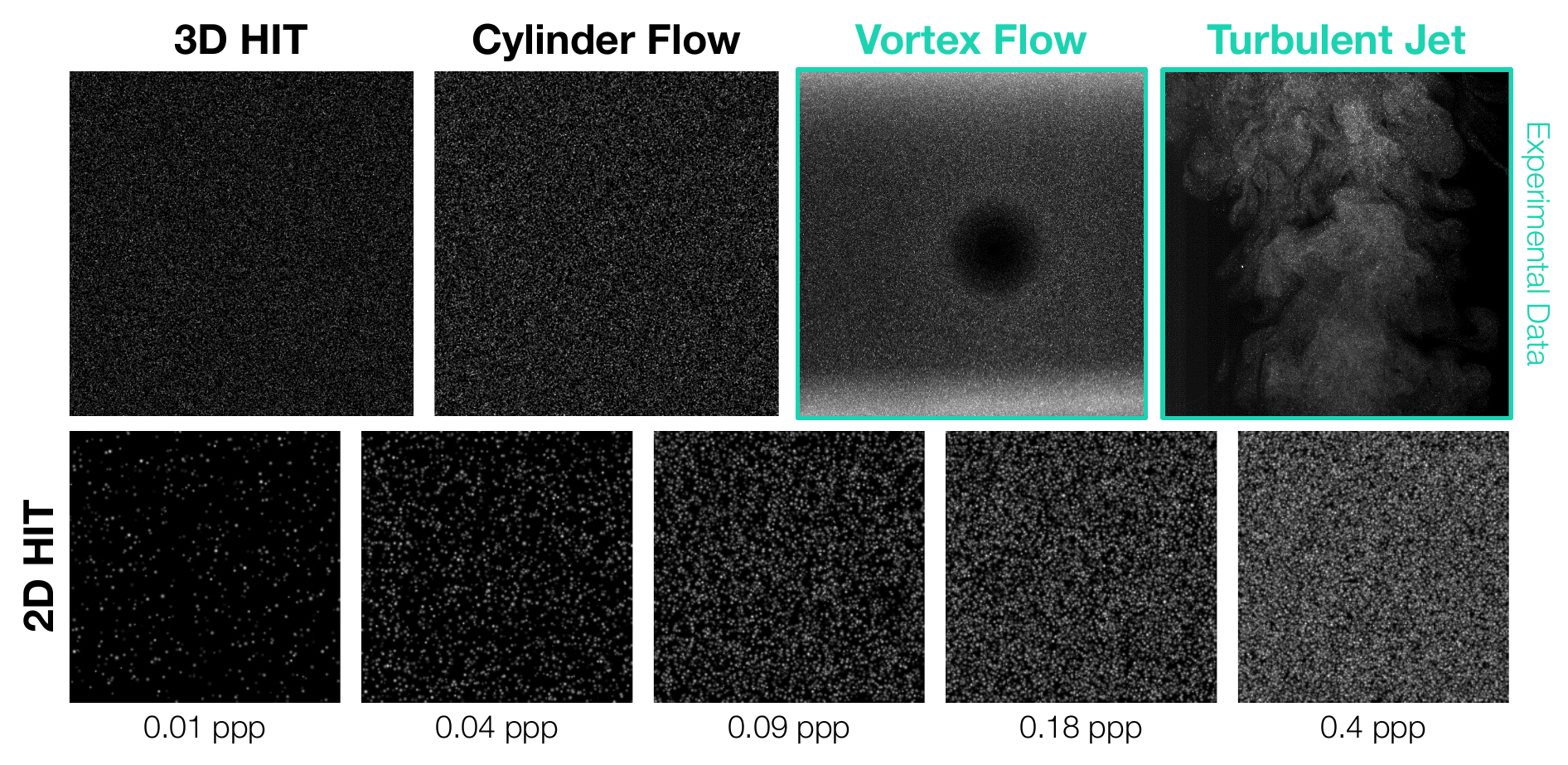}
    \caption{Sample particle images from synthetic and experimental flow cases. Experimental cases from prior PIV challenges \cite{Stanislas2003, Stanislas2008} are outlined in teal.}
    \vspace*{-0.66em}
    \label{fig:cases}
\end{figure*}

We evaluate NOF across a range of synthetic and experimental scenarios. The synthetic tests provide noise-free conditions to assess accuracy under varying seeding densities and examine the effects of out-of-plane motion using both 2D and 3D flows. The experimental cases introduce real-world challenges, including noise, shot-to-shot laser intensity fluctuations, and non-uniform seeding and illumination. In both settings, we benchmark NOF against CC and WOF, which are two widely used, non-data-driven PIV methods. Additionally, we examine the limitations of two data-driven methods, PIV-DCNN \cite{Lee2017} and PIV-LiteFlowNet \cite{Cai2019a}, using selected synthetic cases. Sample images for each scenario are shown in Fig.~\ref{fig:cases}, with detailed descriptions provided in the sections that follow.\par

\subsection{Synthetic Cases}
\label{sec:cases:synthetic}
We use three DNS datasets for synthetic testing: 2D homogeneous isotropic turbulence (HIT) from Carlier and Wieneke \cite{Carlier2005}, 3D HIT from the Johns Hopkins Turbulence Database \cite{Perlman2007}, and a 3D cylinder wake flow from Raissi et al. \cite{Raissi2019}. These cases, referred to as 2D HIT, 3D HIT, and cylinder flow, contain 100, 51, and 201 frames, respectively.\par

To mimic realistic flow imaging conditions, we dimensionalize each dataset under the assumption that the fluid is water. The hypothetical setups are designed to conform to typical PIV configurations. Table~\ref{tab:cases} summarizes key dimensionless and dimensional quantities for each scenario, with Reynolds numbers defined using the energy injection scale (2D and 3D HIT) or the cylinder diameter.\par

\begin{table}[ht]
    \caption{Summary of Parameters and Domain Dimensions for DNS Flow Cases}
    \centering
    \tabcolsep=0.2cm
    \vspace*{.1em}
    \begin{tabular}{c c c c c c c}
        \hline\hline \\[-.75em]
        \multirow{2}{*}{\bf Flow case} & \multirow{2}{*}{\bf $\boldsymbol{Re}$} & \multicolumn{2}{c}{\bf Domain size} & \multicolumn{2}{c}{\bf Time step} & \multirow{2}{*}{\bf Grid}\\ 
        \multicolumn{1}{c}{} & \multicolumn{1}{c}{} & \multicolumn{1}{c}{DNS} & \multicolumn{1}{c}{dim., mm} & \multicolumn{1}{c}{DNS} & \multicolumn{1}{c}{dim., ms}\\
        \hline\\[-.75em]
        \multirow{1}{*}{}  2D HIT & $3000$ & $2\pi \times 2\pi $ & $100 \times 100$ & $0.01$ & $0.845$ & $256 \times 256$ \\
        \multirow{1}{*}{}  3D HIT & $5405$ & $2\pi \times 2\pi \times 5\pi/64 $ & $150 \times 150 \times 5.9 $ & $0.002$ & $0.211$ & $1024 \times 1024 \times 40$ \\
        \multirow{1}{*}{}  Cylinder & $100$ & $5 \times 5 \times 5$ & $5 \times 5$ & $0.08$ & $0.80$ & $51 \times 51 \times 51$ \\
        \hline\hline
    \end{tabular}
    \label{tab:cases}
\end{table}

Generating synthetic particle images involves numerically advecting tracer particles and projecting their position into sensor space. As outlined in \ref{app:camera}, we assume a pinhole camera with a magnification of 1/3. Object distances are set to 120, 150, and 60~mm for the 2D HIT, 3D HIT, and cylinder cases, respectively, with focal lengths of 40, 50, and 20~mm. For planar setups, the image plane is coincident with the physical $x_1{-}x_2$ plane and the camera is centered at $x_1 = x_2 = 0$~mm. The inter-lens spacing is set to 75~mm in our stereo 3D HIT test. In all scenarios, the synthetic laser sheet has a Gaussian profile that straddles the $x_1{-}x_2$ plane about $x_3 = 0$~mm, with a standard deviation in the $x_3$-direction of 0.63~mm.\par

Particles are seeded uniformly throughout the physical domain and advected according to the velocity field. For the 3D HIT and cylinder flows, we employ a second-order Runge--Kutta scheme, while for 2D HIT, we use a forward Euler scheme with at least 50 steps per frame. Periodic boundary conditions are applied to the HIT cases. For the cylinder flow, particles exiting the domain are re-injected at the opposite boundary at a randomized height. Particle locations are projected onto the image plane using Eq.~\eqref{equ:planar transform:forward} for planar setups and Eq.~\eqref{equ:forward transform stereo trans} for stereo setups. Image resolutions are matched to the grid sizes listed in Table~\ref{tab:cases}, except for the cylinder case where images of $841 \times 841$ px are synthesized. Particles are modeled as Gaussian blobs with a standard deviation of 1.25~pixels, giving them an apparent diameter of 3--4 pixels. For 2D HIT, the particle density is varied from 0.01 to 0.4~ppp to assess the effects of seeding density.\par

A key advantage of synthetic testing is the availability of ground truth data, allowing for precise error quantification. For our analysis, we report normalized root-mean-square errors (NRMSEs) of the fluctuating components. This metric allows us to assess whether each technique can capture unsteady flow behavior, as the mean components are relatively easy to estimate. For a quantity of interest, $g$, we apply a Reynolds decomposition, $g = \overline{g} + g^\prime$. The NRMSE for $g$ is
\begin{equation}
    e_g = \frac{\left\langle \left\lVert g^\prime - g_\mathrm{exact}^\prime \right\rVert_2 \right\rangle} {\left\langle \left\lVert g_\mathrm{exact}^\prime \right\rVert_2 \right\rangle}
    \label{equ:NRMSE}    
\end{equation}
where $g_\mathrm{exact}^\prime$ is the exact fluctuating component and $\left\langle \cdot \right\rangle$ indicates a spatial average over the laser sheet,
\begin{equation}
    \left\langle g \right\rangle = \frac{1}{\left| \mathcal{P} \right|} \int_\mathcal{P} g\mathopen{}\left(\mathbf{x}\right) \mathrm{d}\mathbf{x}.
        \label{equ:average}    
\end{equation}
In our Reynolds decomposition, $\overline{g}$ is the long-run time average of $g$.\par

\subsection{Experimental Cases}
\label{sec:cases:experiment}
We also demonstrate NOF on two experimental datasets: a strong vortex flow from the First International PIV Challenge \cite{Stanislas2003} and a time-resolved jet flow from the third challenge \cite{Stanislas2008}. These cases are referred to as vortex flow and turbulent jet flow, respectively.\par

The vortex flow dataset \cite{Stanislas2003} consists of two consecutive frames capturing a vortex formed behind the wingtip of a transport aircraft (ALVAST half-model) in a landing configuration. The main air flow is perpendicular to the laser plane, with a velocity of 60~m~s$^{-1}$, and the measurement plane is positioned 1.64~m behind the wingtip. Images are taken with a progressive scan camera featuring a $1024 \times 1280$~px chip and $6.7 \times 6.7$~$\upmu$m$^2$ pixels. We crop the images to $1024 \times 1024$~px, reducing the field of view from $140 \times 170$~mm$^2$ to $140 \times 140$~mm$^2$. The magnification is approximately 0.05. Since the frame rate is not specified, we follow the original PIV challenge paper and report results in pixels-per-frame (ppf). This dataset is characterized by strong intensity gradients, a non-uniform particle density, and particle size variations, which pose significant challenges for PIV algorithms, making it an excellent test of NOF's robustness.\par

The turbulent jet flow dataset \cite{Stanislas2008} features a nitrogen jet with an exit velocity of approximately 30~m~s$^{-1}$. The nozzle has a diameter of 5~mm, and the flow is seeded with oil-mist particles averaging 5~$\upmu$m in diameter. The domain is illuminated with 527~nm light from a high-speed dual-head PIV laser (New Wave Pegasus) operating at 5~kHz and imaged with a high-speed camera (Photron APX-RS). The camera has a $512 \times 512$~px sensor, $17 \times 17$ $\upmu$m$^2$ pixels, and magnification of 0.3 for the PIV plane. Imaging is performed at 10~kHz using frame straddling to capture the output of both laser heads, which have an inter-frame time separation of 10~$\upmu$s. This dataset was selected to evaluate NOF’s ability to handle noise and inter-frame intensity variations, which are common complications in PIV experiments.\par

\section{Synthetic Results}
\label{sec:results}
This section evaluates NOF performance across multiple synthetic flow cases, benchmarking it against CC, WOF, PIV-DCNN, and PIV-LiteFlowNet-en. For CC, we use a standard multi-pass scheme with $64^2$-, $32^2$-, and $16^2$-pixel interrogation windows and an aggressive 75\% overlap. WOF hyperparameters are \emph{optimally} tuned for each case to ensure a fair comparison. PIV-DCNN and PIV-LiteFlowNet-en are evaluated using publicly available pre-trained models from their public repositories. We explore several NOF variants with distinct regularization schemes and flow parameterizations, first testing their performance on the 2D HIT case under different seeding densities. Spectral error analysis provides additional insights into how well each method captures different flow scales. Next, we demonstrate NOF's accuracy and flexibility with both planar and stereo PIV configurations on 3D HIT data, where strong out-of-plane motion is present. Finally, we showcase NOF's ability to recover pressure fields directly from planar PIV measurements in the cylinder flow case.\par

\subsection{2D HIT}
\label{sec:results:2D HIT}

\subsubsection{Spatial Accuracy}
\label{sec:results:2D HIT:spatial accuracy}
\begin{figure}[ht!]
    \centering\vspace*{-0.66em}
    \includegraphics[width=.9\textwidth]{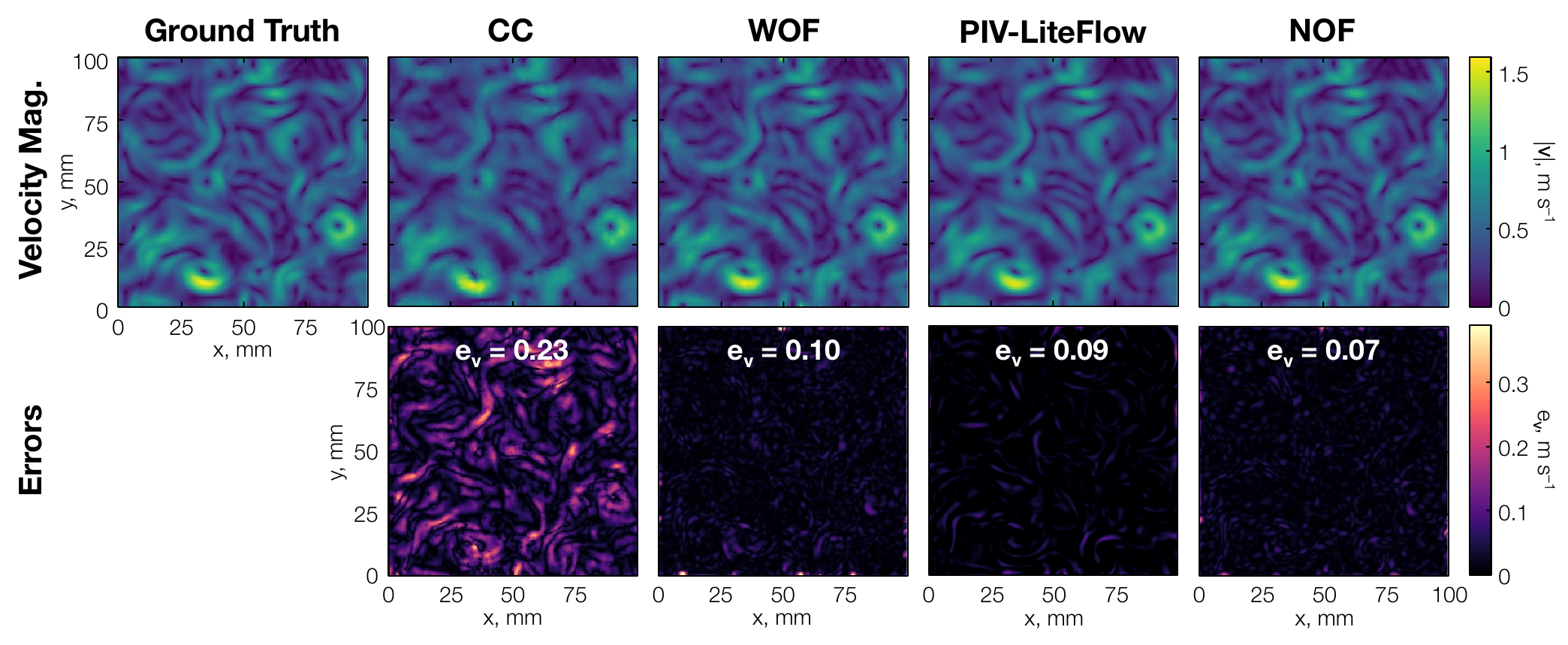}
    \caption{Reconstructed vs. exact velocity magnitude fields for 2D HIT (top) and corresponding absolute error fields (bottom). NRMSEs are superimposed on the error plots.}
    \vspace*{-0.66em}
    \label{fig:2D HIT panel}
\end{figure}

\begin{figure}[ht!]
    \centering\vspace*{-0.66em}
    \includegraphics[width=.75\textwidth]{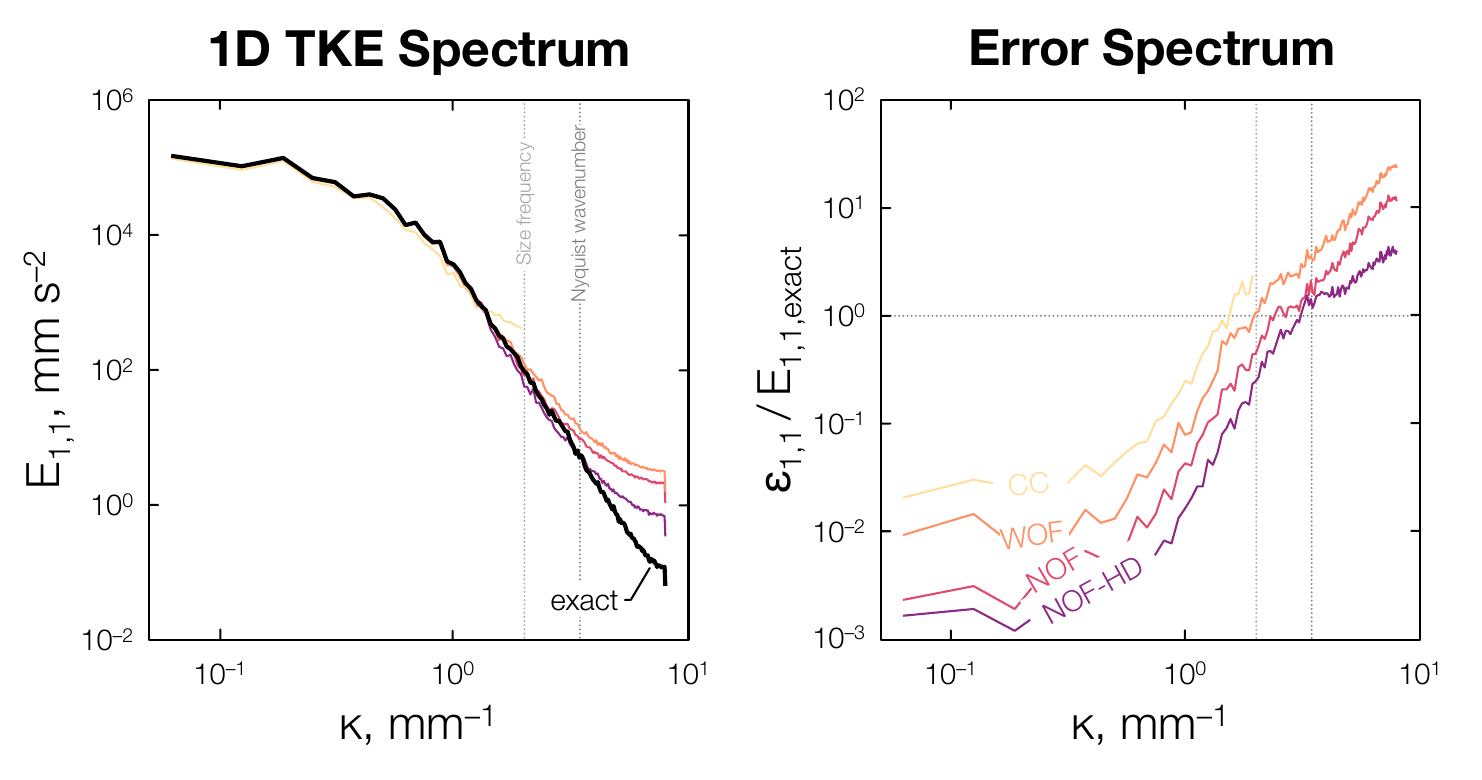}
    \caption{Comparison of 1D TKE spectra (left) and normalized error spectra (right) for 2D HIT reconstructions by various methods. The DNS spectrum is shown for reference.}
    \vspace*{-0.66em}
    \label{fig:2D HIT spectra}
\end{figure}

We begin with the results of the 2D HIT test. Figure~\ref{fig:2D HIT panel} compares a snapshot of the ground truth DNS data to the velocity magnitude fields reconstructed using CC, WOF, PIV-LiteFlowNet-en \cite{Cai2019a}, and NOF. These NOF reconstructions were obtained using only a data loss term, with implicit regularization arising from the network architecture and training algorithm, as described in Sec.~\ref{sec:method:physics-inspired}. The results in Fig.~\ref{fig:2D HIT panel} correspond to a seeding density of 0.17~ppp, which is typical for PIV experiments. This density is sufficient to resolve the Taylor microscale, as indicated in Fig.~\ref{fig:2D HIT spectra}, while also avoiding excessive overlapping of particles, which becomes pronounced at higher seeding densities, such as the 0.4~ppp case shown in Fig.~\ref{fig:cases}.\par

Qualitatively, both NOF and WOF closely match the ground truth velocity field, while CC introduces visible low-pass filtering. This trend is confirmed by the absolute error maps in the bottom row of Fig.~\ref{fig:2D HIT panel}, where WOF and NOF show markedly lower errors than CC. WOF exhibits slightly more artifacts than NOF, particularly near boundaries: an observation consistent with the experimental jet results shown in Fig.~\ref{fig:jet panel} (Sec.~\ref{sec:exp results:turbulent jet flow}). Quantitatively, the NRMSEs in velocity magnitude are 22.8\% for CC, 9.5\% for WOF, 8.6\% for PIV-LiteFlowNet-en, and 6.7\% for NOF. We also tested PIV-DCNN \cite{Lee2017}, which performed poorly with an NRMSE of 33.1\% and is therefore omitted from the figure. While PIV-LiteFlowNet-en performs competitively in this case, comparable to WOF and NOF, subsequent analyses in Secs.~\ref{sec:results:3D HIT} and \ref{sec:results:cylinder flow} highlight the limited generalizability of purely data-driven methods. Accordingly, this work focuses on comparing the three broadly applicable, non-data-driven approaches: CC, WOF, and NOF.\par

\subsubsection{Spectral Analysis}
\label{sec:results:2D HIT:spectral}
While NRMSEs provide a global measure of reconstruction accuracy, spectral analysis reveals how well each method captures flow structures at different length scales. The left side of Fig.~\ref{fig:2D HIT spectra} presents the 1D turbulent kinetic energy (TKE) spectra, $E_{1,1}$, for velocity fields reconstructed by CC, WOF, NOF, and NOF-HD, along with the ground truth spectrum. These spectra are defined as
\begin{equation}
    E_{i,i}\mathopen{}\left(\kappa\right) = \int_0^{2\pi} \left\lvert\hat{v}_{i}\mathopen{}\left(\kappa_1, \kappa_2\right)\right\rvert^2 \kappa \,\mathrm{d}\theta,
\end{equation}
where $\hat{v}_{i}$ is the 2D Fourier transform of $v_i$, $\kappa$ is the wavenumber, $\kappa_1 = \kappa \cos(\theta)$, $\kappa_2 = \kappa \sin(\theta)$, and $\theta$ is the angular coordinate in Fourier space. The spectra indicate the distribution of kinetic energy across wavenumbers. However, a reconstruction may match the energy spectrum of a flow while misrepresenting the underlying velocity field. This is evident in HIT, for example, where distinct snapshots share similar spectra despite differences in their spatial structures. The right side of Fig.~\ref{fig:2D HIT spectra} is a plot of normalized error spectra, quantifying reconstruction accuracy as a function of $\kappa$.\footnote{The sharp drop at the far right of the spectrum (e.g., in Fig.~\ref{fig:2D HIT spectra}) is a numerical artifact introduced by MATLAB's \texttt{pspectrum} function.} Both the standard NOF and NOF-HD variants are analyzed, where NOF-HD embeds mass continuity as a hard constraint (Sec.~\ref{sec:method:physics-based}).\par

To interpret trends in the error spectra, we define particle size and Nyquist-spacing wavenumbers,
\begin{equation}
    \label{equ:wavenumbers}
    \kappa_\mathrm{p} = \frac{2\pi}{d_\mathrm{p}}
    \quad\text{and}\quad
    \kappa_\mathrm{N} = \frac{\pi}{\ell_\mathrm{p}},
\end{equation}
respectively, where $d_\mathrm{p}$ is the apparent particle diameter on the sensor projected into physical space and $\ell_\mathrm{p}$ is the mean distance between particles. Specifically, $\kappa_\mathrm{p}$ represents the spatial wavenumber associated with the particle image size, below which flow features are obscured by the particle. Meanwhile, $\kappa_\mathrm{N}$ indicates the maximum recoverable wavenumber for a given seeding density, following the Shannon--Nyquist sampling theorem \cite{Jerri1977}.\par

We hypothesize that, absent physics-based constraints, the accuracy of OF methods is fundamentally constrained by the lower of these wavenumbers, above which finer flow details cannot be resolved. As shown on the left side of Fig.~\ref{fig:2D HIT spectra}, CC deviates from the ground truth spectrum well before reaching the particle size wavenumber, due to the low cut-off imposed by its large interrogation window. By comparison, WOF and NOF closely match the ground truth spectrum up to the particle size wavenumber. Remarkably, NOF-HD surpasses this limit, maintaining accuracy up to $\kappa_\mathrm{N}$. This highlights the value of incorporating physical constraints into OF algorithms, enabling better resolution of fine-scale flow features.\par

Spectral errors, shown on the right side of Fig.~\ref{fig:2D HIT spectra}, are normalized by the true kinetic energy at each wavenumber, $\varepsilon_{1,1} / E_{1,1,\mathrm{exact}}$. This normalization provides a scale-dependent assessment of reconstruction accuracy. Errors follow a consistent trend across all methods: minimal at small wavenumbers, followed by a rapid increase at higher wavenumbers. Both NOF variants outperform CC and WOF across the spectrum. Notably, at the particle size wavenumber, CC and WOF approach 100\% error, whereas NOF-HD maintains an error close to 20\%. Curiously, while WOF and vanilla NOF exhibit an exponential growth in error at high wavenumbers, NOF-HD shows a more gradual increase, especially beyond the Nyquist wavenumber. This suggests that enforcing mass continuity suppresses particle-induced artifacts and random errors at high wavenumbers. However, NOF-HD was unable to resolve flow behavior at wavenumbers beyond $\kappa_\mathrm{N}$, indicating a resolution limit associated with particle sampling.\par

To evaluate flow reconstruction accuracy across both resolved and unresolved scales, defined with respect to a cut-off wavenumber of $\kappa_\mathrm{N}$, we compute \emph{sub}- and \emph{super}-Nyquist velocity errors. Sub-Nyquist errors are calculated by low-pass filtering the velocity field in the Fourier domain, setting all components with $\kappa \geq \kappa_\mathrm{N}$ to zero. The filtered velocity fields are then mapped back to the physical domain using an inverse Fourier transform, and sub-Nyquist errors are evaluated using Eq.~\eqref{equ:NRMSE}. Super-Nyquist errors are obtained similarly by applying a high-pass filter, setting all components with $\kappa < \kappa_\mathrm{N}$ to zero, and then evaluating errors in the physical domain. This decomposition separates reconstruction accuracy at wavenumbers that are, in principle, directly resolvable from the measurements (sub-Nyquist) from high-wavenumber accuracy, which is primarily influenced by priors (regularization) and the inductive biases of the reconstruction algorithm.\par

Table~\ref{tab:2D HIT errors} provides sub- and super-Nyquist errors for the 2D HIT case reconstructed using CC, WOF, both supervised ML methods, and several NOF variants. The time-resolved NOF variants (NOF-TR and NOF-phys) were trained on a sequence of 20 particle images, with reported errors corresponding to the fifth frame to match the non-time-resolved tests.\par

\begin{table}[ht]
    \caption{Errors for 2D HIT Reconstructions}
    \centering
    \tabcolsep=0.2cm
    \vspace*{.1em}
    \begin{tabular}{c c c c c c c c c}
        \hline\hline \\[-.75em]
        \multirow{3}{*}{} & \multicolumn{8}{c}{\textbf{Algorithm}}\\ 
        & \multicolumn{2}{c}{\textit{Traditional}}& \multicolumn{2}{c}{\textit{Supervised}}&\multicolumn{2}{c}{\textit{Unsup. (inst.)}}&\multicolumn{2}{c}{\textit{Unsup. (time)}}\\
        & \multicolumn{1}{c}{CC} & \multicolumn{1}{c}{WOF} & 
        \multicolumn{1}{c}{PIV-DCNN} & \multicolumn{1}{c}{LiteFlow} & 
        \multicolumn{1}{c}{NOF} & \multicolumn{1}{c}{NOF-HD} & \multicolumn{1}{c}{NOF-TR} & \multicolumn{1}{c}{NOF-phys}\\ 
        \hline 
        \multirow{1}{*}{}  &  &  &  &  &  &  \\[-.75em]
        \multirow{1}{*}{} Sub-Nyq. & $20.94$ & $9.13$ & $29.18$ & $8.48$ & $6.76$ & $\mathbf{5.26}$ & $8.95$ & $8.16$ \\
        \multirow{1}{*}{} Super-Nyq. & N/A & $188.42$ & $\mathbf{42.50}$ & $133.66$ & $134.78$ & $114.12$ & $141.47$ & $106.22$ \\
        \hline\hline
    \end{tabular}
    \label{tab:2D HIT errors}
\end{table}

Among all methods, WOF exhibits the highest super-Nyquist error, likely due to artifacts introduced by wavelet-domain regularization. The impact of NOF variants is also clear. NOF-TR incurs slightly higher sub- and super-Nyquist errors (attributable to the greater sampling and data density demands of its larger spatiotemporal domain) but achieves a 20-fold increase in data compression and reduces per-frame training time from 3--7 minutes to just 20--45 seconds. More notably, NOF-phys improves super-Nyquist accuracy beyond what is achieved with the hard divergence constraint in NOF-HD. While the inclusion of Navier--Stokes residuals only modestly improves sub-Nyquist performance, it significantly suppresses spurious high-frequency artifacts, leading to a marked reduction in super-Nyquist errors. These findings highlight the value of physics-based constraints, particularly for mitigating non-physical fluctuations at high wavenumbers. However, the 2D HIT DNS from Carlier and Wieneke \cite{Carlier2005} sustains turbulence using forcing and dissipation terms that are not included in the momentum equation residuals used in the NOF-phys penalty. This discrepancy may limit the applicability of Eq.~\eqref{equ:physics constraint:NS} for this case.\par

\subsubsection{Effects of Particle Density}
\label{sec:results:2D HIT:particle density}
\begin{figure}[ht!]
    \centering\vspace*{-0.66em}
    \includegraphics[width=.8\textwidth]{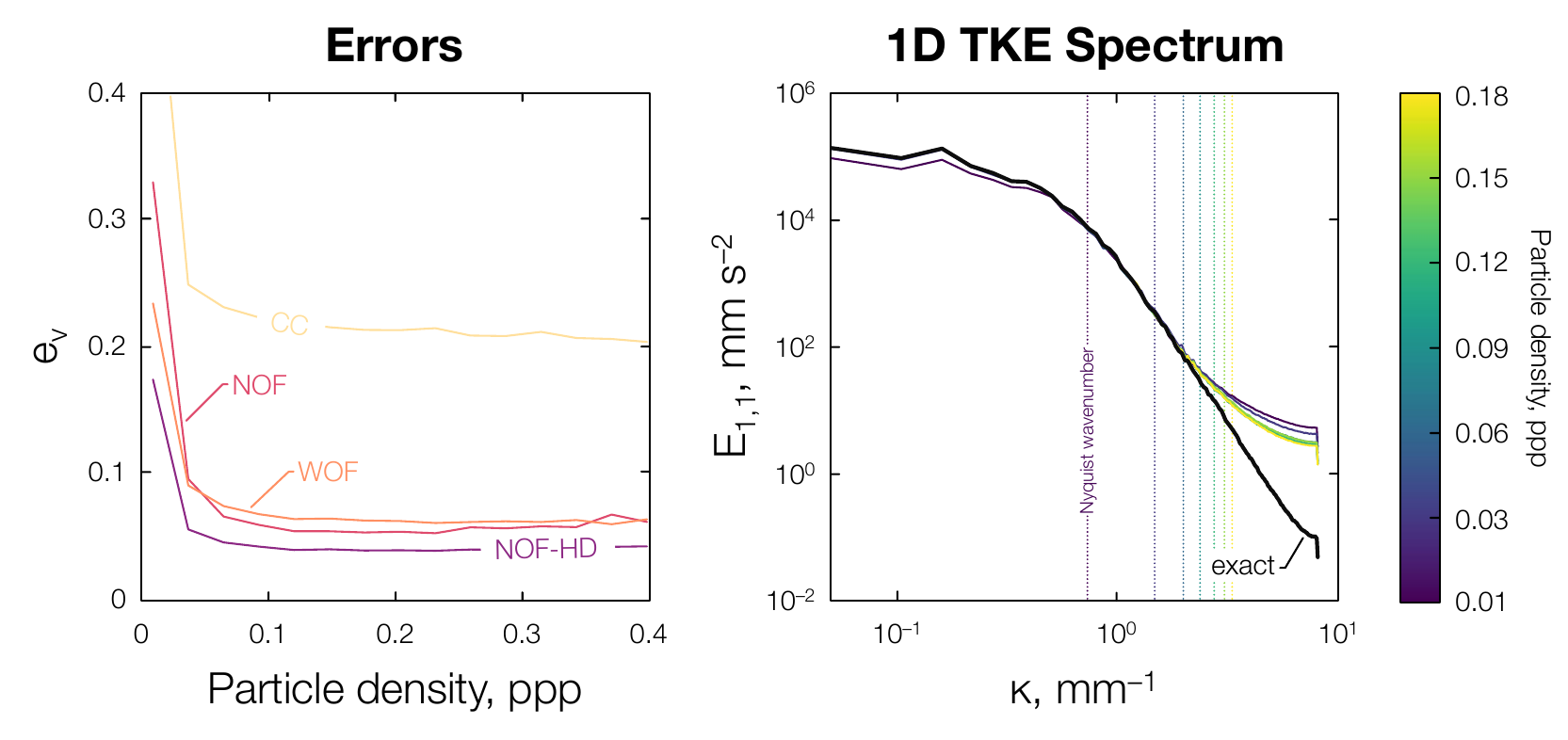}
    \caption{NRMSEs for 2D HIT reconstructions across particle seeding densities (left) and corresponding 1D TKE spectra from NOF (right). Note the close agreement with ground truth at low wavenumbers for ppp $> 0.03$.}
    \vspace*{-0.66em}
    \label{fig:2D HIT ppp}
\end{figure}

In addition to the tests at a fixed seeding density of 0.17~ppp, we evaluated the performance of CC, WOF, NOF, and NOF-HD across varying particle densities, with results shown in Fig.~\ref{fig:2D HIT ppp}. Our analysis focuses on NOF and NOF-HD, as they represent the two core variants with favorable accuracy (see Table~\ref{tab:2D HIT errors}). The left plot in Fig.~\ref{fig:2D HIT ppp} depicts NRMSEs of velocity magnitude fields, averaged over ten consecutive snapshots, across a range of particle densities from 0.01 to 0.4~ppp. Exemplary snapshots at a range of particle densities can be seen in the bottom row of Fig.~\ref{fig:cases}.\par

Once again, NOF outperforms WOF at all but the lowest seeding densities and consistently surpasses CC across the entire range. NOF-HD delivers the best overall performance, underscoring the benefits of imposing hard physical constraints for flow reconstruction. The right plot in Fig.~\ref{fig:2D HIT ppp} shows the 1D TKE spectra of NOF-reconstructed fields at various seeding densities. As expected, reducing the density diminishes low-wavenumber content while amplifying high-frequency noise. This effect arises because increased inter-particle spacing allows more spurious velocity oscillations between particles. Such oscillations are not penalized by the loss function when the particle spacing grows too large.\par

The plot also includes vertical lines marking the Nyquist wavenumber for each seeding density, which roughly aligns with the point at which the spectra diverge for ppp values below 0.06. Further increases in seeding density do not yield significant improvements in resolving finer flow scales, as the finite size of particles in the images becomes the limiting factor.\par

\subsection{Stereo 3D HIT}
\label{sec:results:3D HIT}
\begin{figure}[ht!]
    \centering\vspace*{-0.66em}
    \includegraphics[width=.9\textwidth]{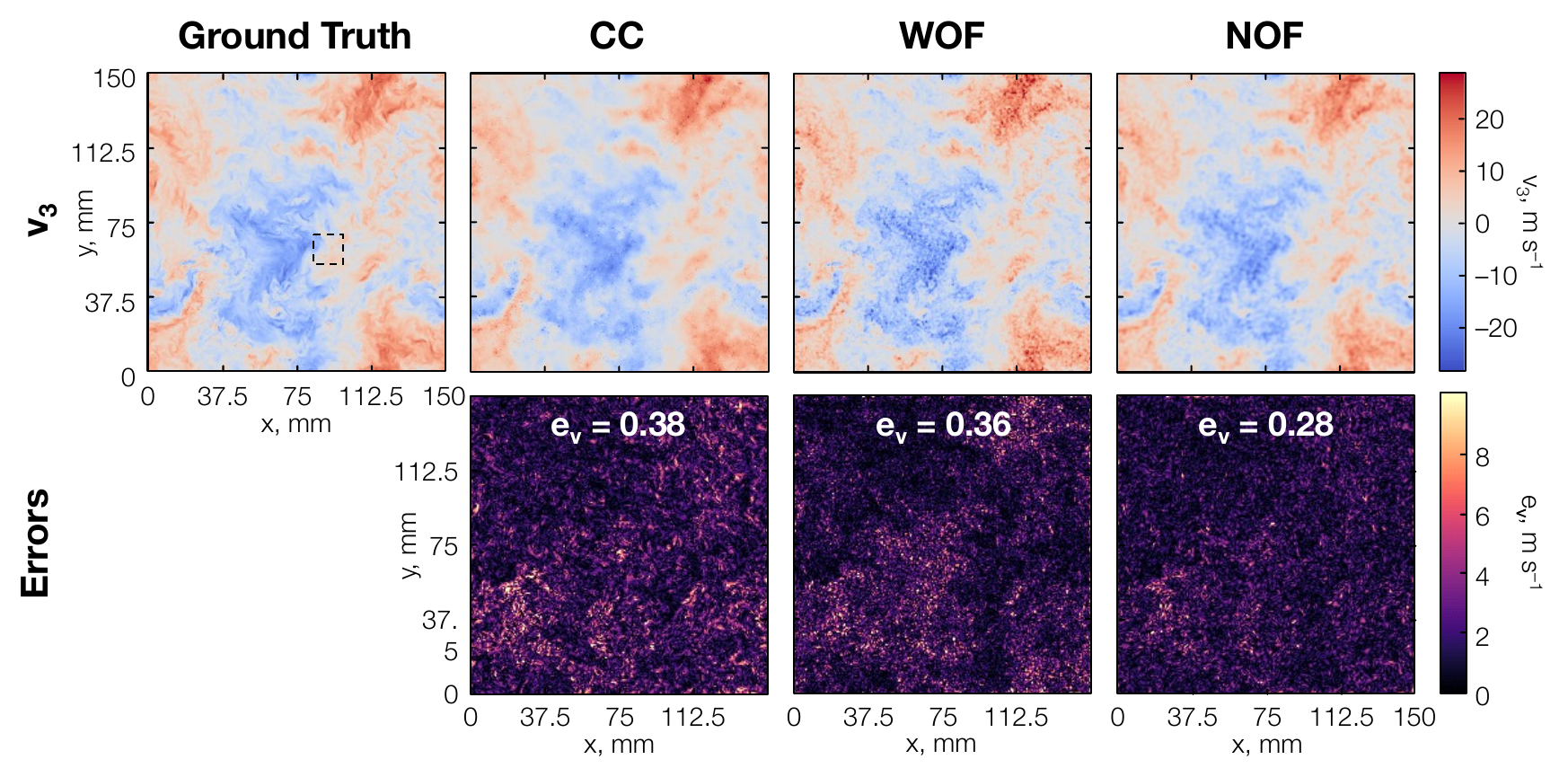}
    \caption{Reconstructed vs. exact velocity magnitude fields for 3D HIT (top) and corresponding absolute error fields (bottom) for a stereo setup. NRMSE values are superimposed.}
    \vspace*{-0.66em}
    \label{fig:3D HIT stereo panel}
\end{figure}

In this section, we compare data processing methods for stereo PIV. To reduce the computational cost of NOF, we use a forward Euler step size of $\Delta t$ to solve Eq.~\eqref{equ:NOF:displacement transform:advection}, instead of $\Delta t/5$, which is sufficient for the 3D HIT dataset due to its small, nearly linear displacements. Unlike conventional stereo PIV methods that independently reconstruct two 2D2C fields before combining them into a single 2D3C velocity field, stereo NOF directly reconstructs a 2D3C velocity field by simultaneously minimizing data losses for both cameras. Figure~\ref{fig:3D HIT stereo panel} compares out-of-plane velocity fields reconstructed by CC, WOF, and stereo NOF against the ground truth DNS. NOF achieves the highest accuracy, with a NRMSE of 28.4\%, outperforming CC (37.5\%), WOF (35.9\%), and the supervised learning methods PIV-DCNN (33.2\%) and PIV-LiteFlowNet-en (75.4\%), which are omitted from the figure for brevity.\par

\begin{table}[ht]
    \caption{Errors for 3D HIT Reconstructions}
    \centering
    \tabcolsep=0.2cm
    \vspace*{.1em}
    \begin{tabular}{c c c c c c c c c}
        \hline\hline \\[-.75em]
        \multirow{3}{*}{\textbf{Component}} & \multicolumn{6}{c}{\textbf{Algorithm}} \\ 
        & \multicolumn{2}{c}{\textit{Traditional}}& \multicolumn{2}{c}{\textit{Supervised}}&\multicolumn{2}{c}{\textit{Unsupervised}}\\
        & \multicolumn{1}{c}{CC} & \multicolumn{1}{c}{WOF} & \multicolumn{1}{c}{PIV-DCNN} & \multicolumn{1}{c}{LiteFlow} &\multicolumn{1}{c}{NOF-recomb} & \multicolumn{1}{c}{stereo NOF}\\
        \hline\\[-.75em]
        \multirow{1}{*}{} $v_1$ & $20.24$ & $21.41$ & $28.82$ & $38.35$ & $20.11$ & $\mathbf{19.44}$ \\
        \multirow{1}{*}{} $v_2$ & $27.40$ & $21.00$ & $28.26$ & $39.56$ & $20.08$ & $\mathbf{19.59}$ \\
        \multirow{1}{*}{} $v_3$ & $37.46$ & $35.91$ & $33.24$ & $75.43$ & $31.40$ & $\mathbf{28.42}$ \\[.15em]
        \multirow{1}{*}{} Sub-Nyq. & $29.20$ & $23.73$ & $26.58$ & $45.59$& $22.57$ & $\mathbf{20.91}$\\
        \multirow{1}{*}{} Super-Nyq. & $129.2$ & $143.5$ & $\mathbf{93.60}$ & $94.75$& $104.7$ & $101.4$\\
        \hline\hline
    \end{tabular}
    \label{tab:3D HIT errors}
\end{table}

As shown in Fig.~\ref{fig:3D HIT stereo panel}, NOF significantly reduces out-of-plane errors, highlighting its advantage for stereo PIV. Table~\ref{tab:3D HIT errors} presents errors for the 3D HIT case, including NRMSEs for each velocity component, $\mathbf{v} = (v_1, v_2, v_3)$. In addition to stereo NOF, results are shown for ``NOF-recomb,'' where vanilla NOF is applied separately to each view before recombination, as in CC and WOF. While in-plane errors are comparable across methods, accuracy improves progressively from CC to WOF to NOF-recomb to stereo NOF, with the greatest reduction in error observed for the out-of-plane component, $v_3$. The improvement from WOF to NOF-recomb reflects the benefits of NOF's neural parameterization, while the additional gains from NOF-recomb to stereo NOF demonstrate the advantages of a unified reconstruction, which eliminates errors introduced during the recombination step.\par

\begin{figure}[ht!]
    \centering\vspace*{-0.66em}
    \includegraphics[width=0.9\textwidth]{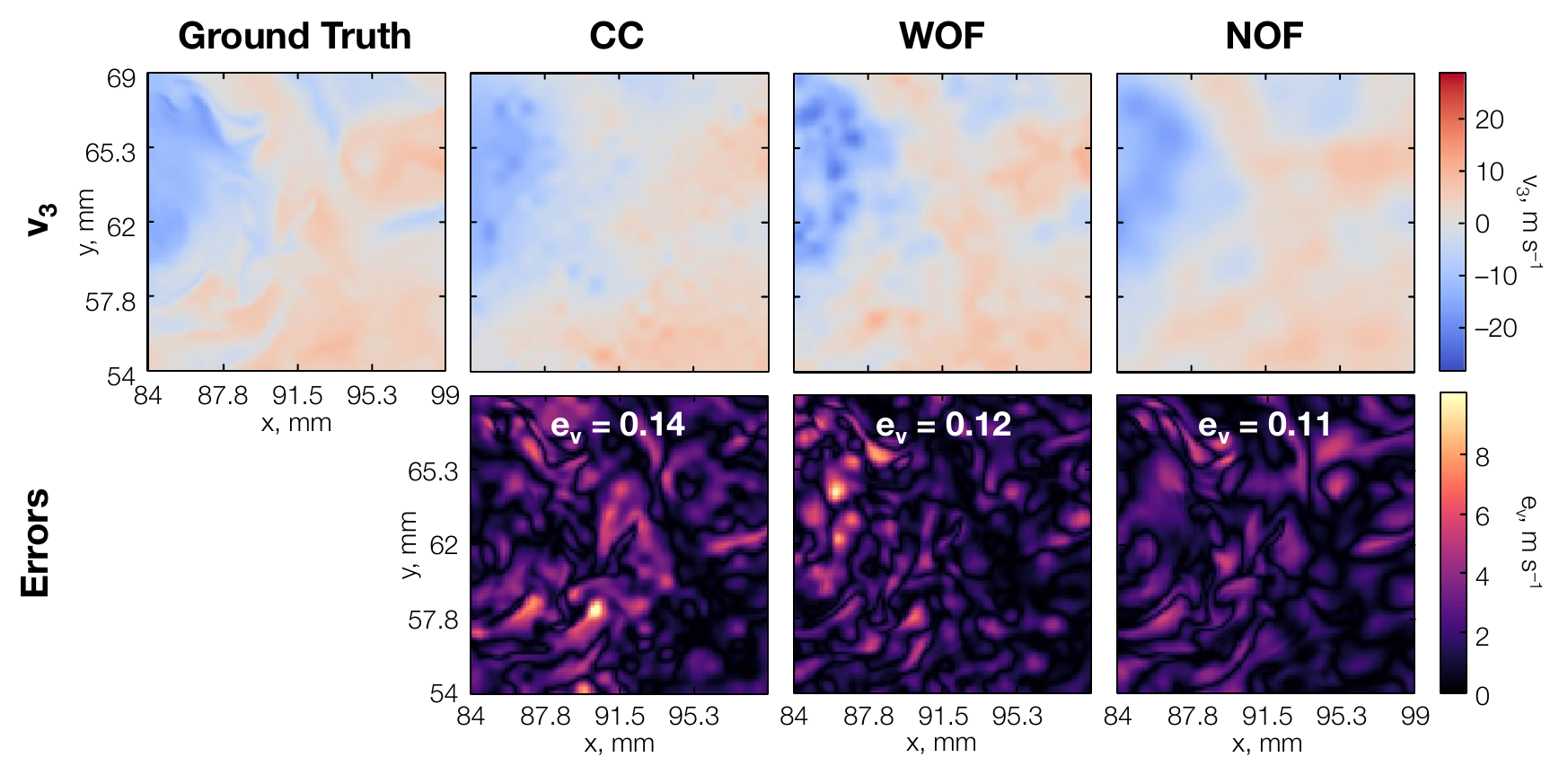}
    \caption{Zoomed views of the velocity magnitude fields outlined in Fig.~\ref{fig:3D HIT stereo panel} (top) and the corresponding absolute error fields (bottom).}
    \vspace*{-0.66em}
    \label{fig:3D HIT stereo panel zoom}
\end{figure}

Although Table~\ref{tab:3D HIT errors} shows lower quantitative errors for NOF compared to WOF, Fig.~\ref{fig:3D HIT stereo panel} may suggest that WOF resolves finer flow features, which is counterintuitive. However, many of these details are spurious artifacts rather than physically accurate structures. Figure~\ref{fig:3D HIT stereo panel zoom} provides a close-up view of the region indicated by the dotted box in Fig.~\ref{fig:3D HIT stereo panel}, revealing artifacts in the WOF estimate, particularly near the left boundary. These amplified errors occur in regions with strong out-of-plane motion, further illustrating how inaccuracies in $v_3$ propagate into in-plane velocity estimates, degrading overall reconstruction fidelity.\par

\begin{figure}[ht!]
    \centering\vspace*{-0.66em}
    \includegraphics[width=0.95\textwidth]{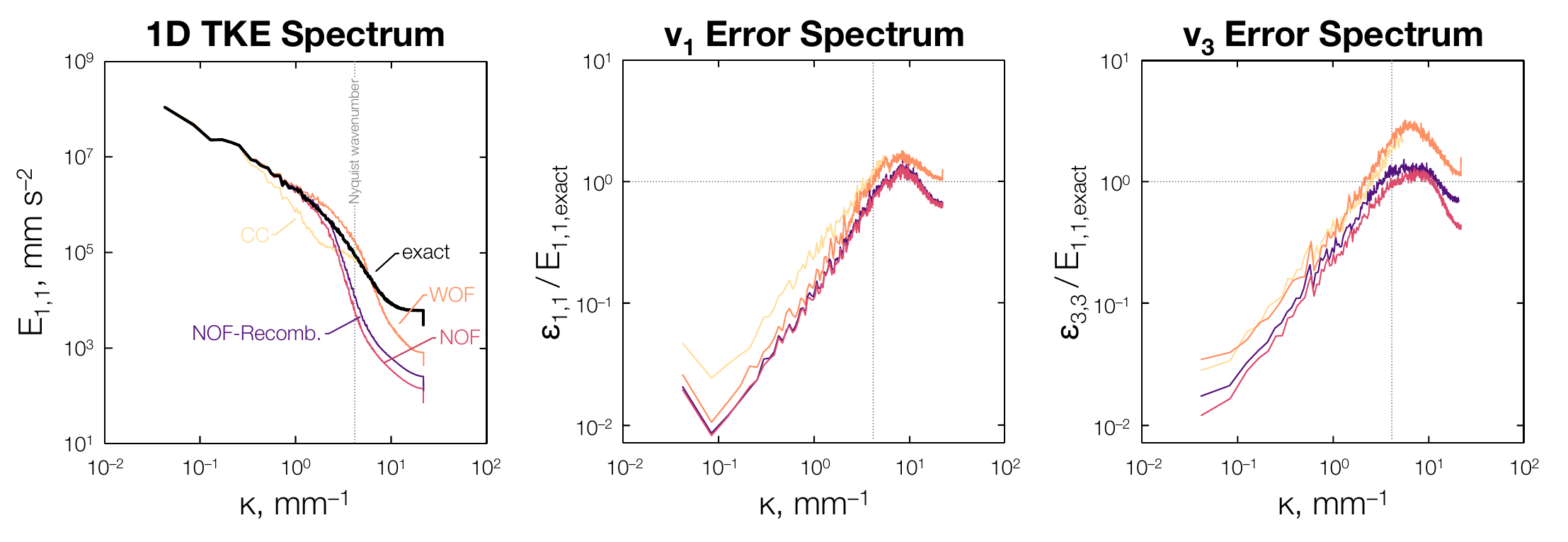}
    \caption{1D TKE spectrum of stereo reconstructions for 3D HIT compared to DNS (left), with normalized error spectra for $v_1$ (middle) and $v_3$ (right).}
    \vspace*{-0.66em}
    \label{fig:3D HIT stereo spectra}
\end{figure}

Figure~\ref{fig:3D HIT stereo spectra} quantifies spectral content through 1D TKE and normalized error spectra for the horizontal and out-of-plane velocity components, $v_1$ and $v_3$. Since the spectra for both in-plane components, $v_1$ and $v_2$, are nearly identical, results for $v_2$ are omitted. This figure compares CC, WOF, NOF-recomb, and stereo NOF. The normalized error spectra reveal a clear, stepwise reduction in error from CC to WOF to NOF-recomb to stereo NOF. For the in-plane components, error reductions are most pronounced at high wavenumbers, indicating a reduction of spurious, high-frequency artifacts. The most significant improvement is seen in the out-of-plane component, $v_3$, where NOF consistently reduces errors across the entire wavenumber range.\par

Table~\ref{tab:3D HIT errors} reports sub- and super-Nyquist errors for the 3D HIT dataset, computed as described in Sec.~\ref{sec:results:2D HIT:spectral}. Stereo NOF and NOF-recomb exhibit nearly 100\% error in the super-Nyquist region, whereas CC and WOF show even higher errors. This discrepancy arises because NOF fields contain minimal high-frequency content, while WOF introduces more spurious details, as seen in the leftmost plot of Fig.~\ref{fig:3D HIT stereo spectra}. Unlike the traditional and unsupervised methods, both supervised ML algorithms have low super-Nyquist errors, likely due to the inherent damping of high-frequency content. Notably, stereo NOF outperforms NOF-recomb across both spectral regions, reinforcing the advantage of a unified reconstruction approach, even when controlling for other factors. These results highlight NOF's superior ability to capture out-of-plane motion accurately, an advantage that becomes especially valuable in complex 3D flow scenarios.\par

A key objective of stereo NOF is to enable pressure recovery from PIV in complex flows. However, in this 3D HIT case, the absence of out-of-plane velocity gradients prevents accurate pressure estimation. Without $\partial{v_3}/\partial{x_3}$ information, the continuity constraint can also become inadequate. The physics-informed NOF variants (NOF-HD and NOF-phys) are thus omitted from this analysis. Both stereo NOF and NOF-recomb diverge from the ground truth spectrum at the same wavenumber as WOF, but their spectral content decays more rapidly beyond this point, likely due to limited expressivity of the networks.\par

\subsection{Cylinder Flow}
\label{sec:results:cylinder flow}
\begin{figure}[ht!]
    \centering\vspace*{-0.66em}
    \includegraphics[width=.9\textwidth]{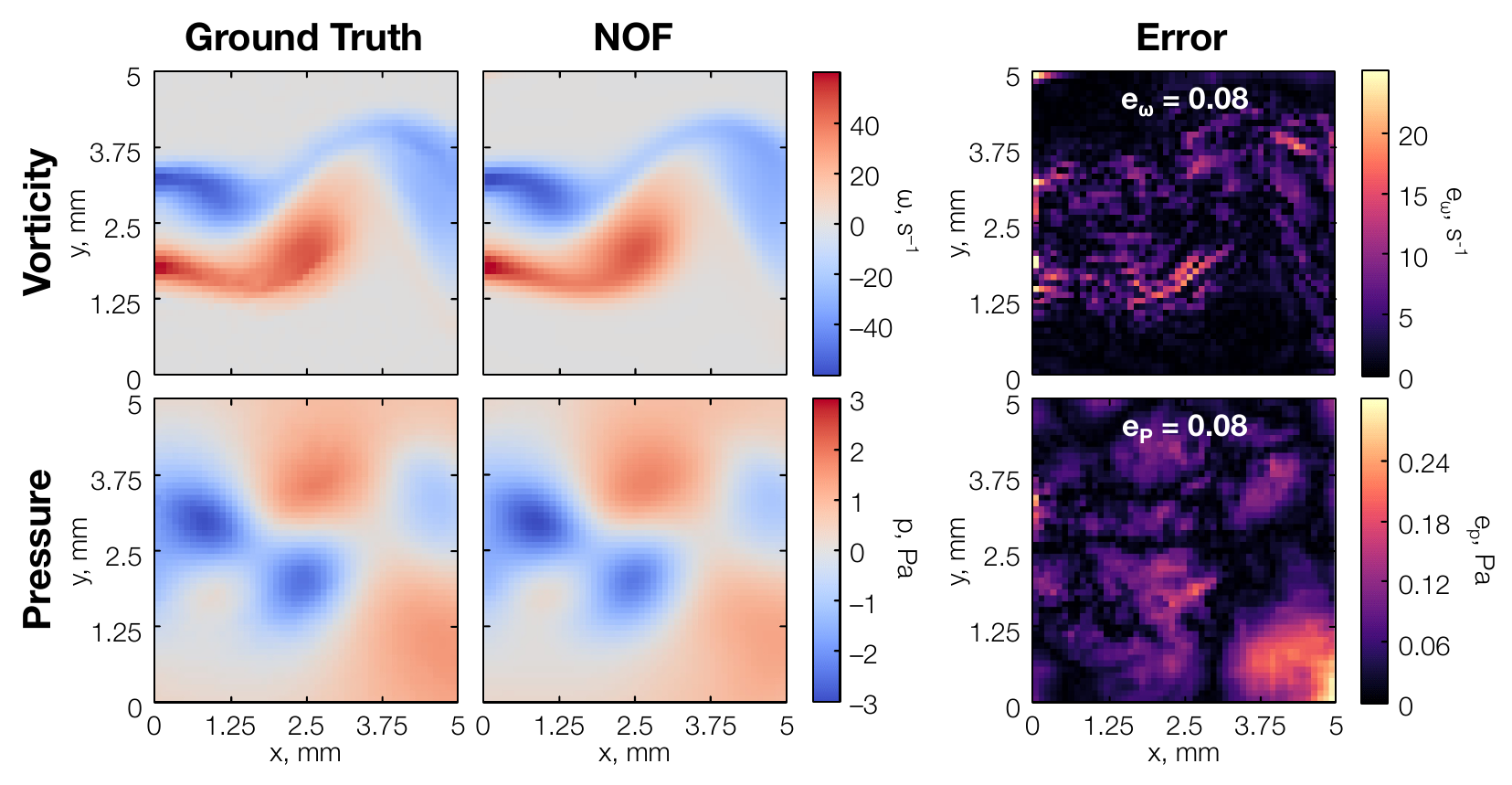}
    \caption{Reconstructed vs. exact vorticity fields (top) and pressure fields (bottom) for the cylinder flow case. Corresponding point-wise error fields shown at right. NRMSE values are superimposed.}
    \vspace*{-0.66em}
    \label{fig:cylinder panel}
\end{figure}

In this section, we present results from the cylinder flow test case, focusing on pressure reconstruction using the NOF-phys variant. Unlike traditional methods that estimate pressure through post-processing, NOF-phys directly infers pressure from PIV images by embedding OF constraints and governing equations into the loss function. Figure~\ref{fig:cylinder panel} compares the reconstructed vorticity and pressure fields to the ground truth. The reconstructions accurately capture dominant features such as vortex shedding, pressure depressions within vortex cores, and quiescent regions with negligible vorticity. Minor smoothing is observed near steep vorticity gradients, but overall accuracy remains high, with NRMSEs of 8.28\% for vorticity and 8.18\% for pressure. Notably, and in contrast to the 2D HIT case (Sec.~\ref{sec:results:2D HIT}), enforcing a hard divergence constraint increases errors to 9.75\% (vorticity) and 11.97\% (pressure). This degradation may stem from out-of-plane effects in this quasi-2D configuration or from optimization challenges associated with enforcing a strictly divergence-free manifold.\par

Although this case emphasizes physics-based pressure reconstruction, we also assessed velocity accuracy in isolation. Vanilla NOF yields the lowest vorticity error (5.87\%), outperforming both WOF (7.83\%) and CC (9.90\%). Supervised methods performed poorly: PIV-DCNN produced an error of 187.86\%, while PIV-LiteFlowNet-en reached 80.55\%. For PIV-DCNN, the majority of this error is concentrated in two large regions of slow-moving flow constituting roughly one-third of the domain. Excluding these regions reduces its error to just 1.8\%, highlighting the method's instability in flows with spatially varying dynamics. PIV-LiteFlowNet-en, meanwhile, produced noisy and inconsistent predictions across the field. These results echo trends observed in the 2D HIT case, reinforcing the difficulty of generalization in supervised models, particularly in flows with anisotropic or non-uniform spatial features. They also suggest a modest trade-off between the accuracy of vorticity and pressure reconstructions, potentially due to the finite expressivity of the network. Future work will investigate how architecture and seeding density affect the performance of joint velocity-pressure reconstruction compared to velocity-only inversion.\par

Although pressure Poisson solvers offer an alternative means of estimating pressure from velocity data, they are highly sensitive to velocity errors, which propagate into the pressure field without any mechanism for correction \cite{Miotto2025}. Regularization is typically necessary to suppress this amplification \cite{Pan2016, Pryce2024}. In contrast, NOF-phys simultaneously reconstructs velocity and pressure fields by minimizing residuals of the Navier--Stokes equations. This joint optimization allows each field to inform the other, promoting consistency with both physical laws and observed image data. In the future, we will directly compare NOF-phys and Poisson-based solvers to evaluate their relative accuracy, stability, and applicability across flow regimes.\par

\section{Experimental Results}
\label{sec:exp results}
This section evaluates the quality of NOF reconstructions in two experimental tests with distinct flow characteristics and realistic measurement challenges. These cases were chosen to span a broad range of flow conditions, testing NOF's performance in both smooth and turbulent regimes. The first case, a vortex flow (Sec.~\ref{sec:exp results:vortex flow}), features a single, smooth vortex in the wake of an airfoil \cite{Stanislas2003}. This provides a controlled test where the expected solution is smooth, allowing us to assess NOF's ability to handle noise while preserving flow structure. The second case, a turbulent jet (Sec.~\ref{sec:exp results:turbulent jet flow}), is a high-Reynolds-number flow with broadband velocity fluctuations, testing NOF's capability to reconstruct fine-scale turbulence from image data with experimental noise.\par

For both cases, CC is performed using a standard three-pass cascade with 64\textsuperscript{2}-, 32\textsuperscript{2}-, and 16\textsuperscript{2}-pixel interrogation windows and 75\% overlap. WOF hyperparameters are tuned for optimal performance based on prior studies.\par

\subsection{Vortex Flow}
\label{sec:exp results:vortex flow}
\begin{figure}[ht!]
    \centering\vspace*{-0.66em}
    \includegraphics[width=.9\textwidth]{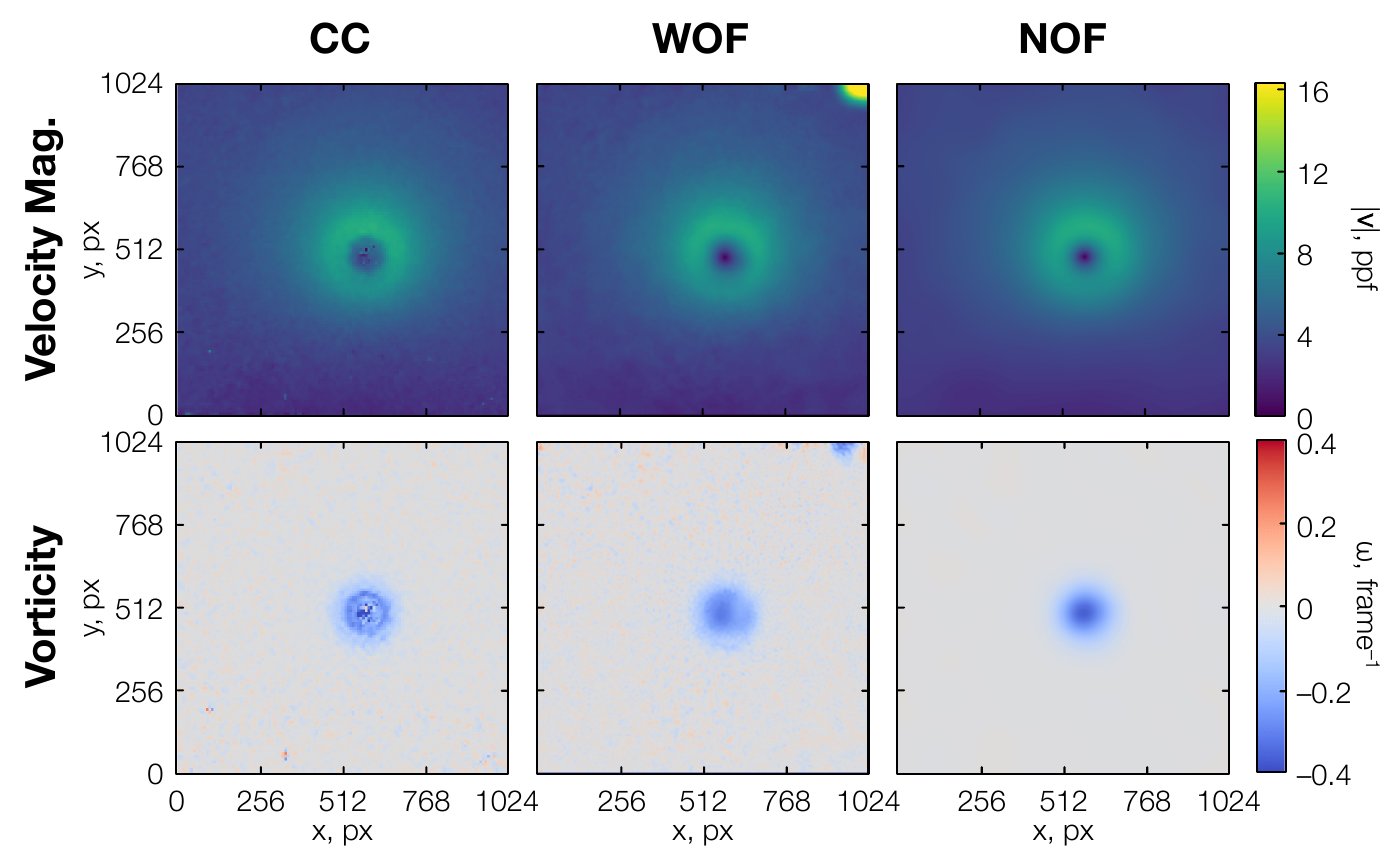}
    \caption{Reconstructed vortex flow velocity magnitude fields (top) and vorticity fields (bottom) for each method.}
    \vspace*{-0.66em}
    \label{fig:vortex panel}
\end{figure}

Figure~\ref{fig:vortex panel} compares the velocity magnitude and vorticity fields reconstructed by CC, WOF, and NOF. While hyperparameter selection for CC and NOF was straightforward, WOF required extensive tuning due to its sensitivity to noise. For WOF, we tested over 30 hyperparameter combinations before selecting what we deemed to be the best result, which still shows artifacts near the corner of the field. Many other combinations produced larger artifacts closer to the vortex core. Both CC and WOF exhibit grainy textures due to image noise, whereas NOF is more resilient to this effect. The vorticity fields highlight this noise sensitivity, particularly near the vortex core, where both CC and WOF produce artifacts that distort the expected vortex shape.\par

\begin{figure}[ht!]
    \centering\vspace*{-0.66em}
    \includegraphics[width=.9\textwidth]{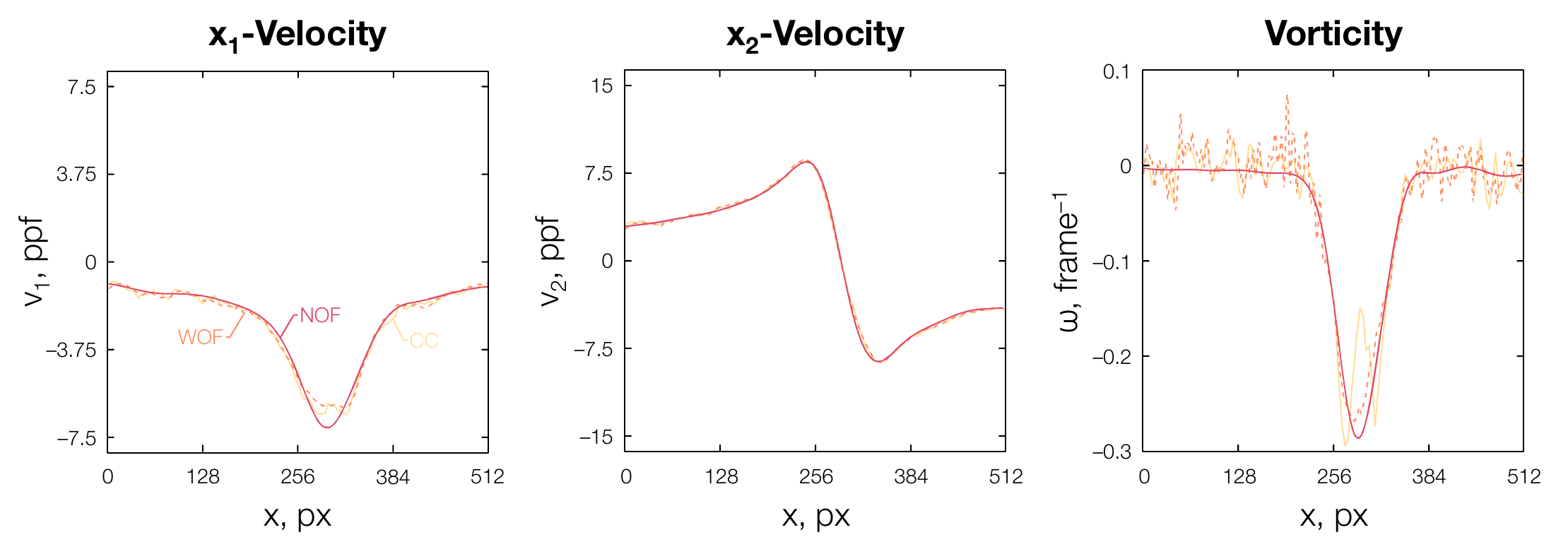}
    \caption{Velocity and vorticity cut plots for the vortex flow case: $v_1$ (left), $v_2$ (middle), and vorticity (right).}
    \vspace*{-0.66em}
    \label{fig:vortex cuts}
\end{figure}

Figure~\ref{fig:vortex cuts} showcases horizontal cut plots of both velocity components and vorticity, taken at a slight offset from the vortex core at $x_2 = 490$~px. The vorticity plot reveals the impact of noise on the reconstructions: while CC and WOF exhibit \emph{a few} similar spikes, most of the oscillations are inconsistent between the two methods, suggesting they are in fact non-physical artifacts. Additionally, both the $v_1$ and vorticity cut plots display clear artifacts within the vortex core. Images for this case show signs of significant tracer thinning in the core, suggesting that the particles are subject to inertial transport rather than passively following the surrounding fluid. Tracer thinning complicates the accurate reconstruction of flow in the core region, and the inertial behavior means the reconstructed fields will not perfectly represent the true fluid velocity field.\par

While it is difficult to conclusively determine whether NOF provides a more accurate reconstruction due to the absence of ground truth data, NOF aligns more closely with the expected shape of a vortex than either CC or WOF. This indicates that NOF may be more resilient to noise than CC or WOF, though further research is needed to quantitatively assess the impact of noise on NOF's performance.\par

\subsection{Turbulent Jet Flow}
\label{sec:exp results:turbulent jet flow}
Unlike the simpler hyperparameter selection problem in the vortex flow case, the jet flow contains significant energy content across a broad range of frequencies, comparable to the 3D HIT field. Therefore, a large Fourier encoding is needed to accurately capture the jet. However, the increased expressivity needed for this representation also makes NOF more prone to noise-related artifacts. To address this, we applied the div--curl regularization penalty from Eq.~\eqref{equ:reg:div-curl}, which mitigates noise artifacts while preserving the network's ability to represent the flow accurately.\par

\begin{figure}[ht!]
    \centering\vspace*{-0.66em}
    \includegraphics[width=.9\textwidth]{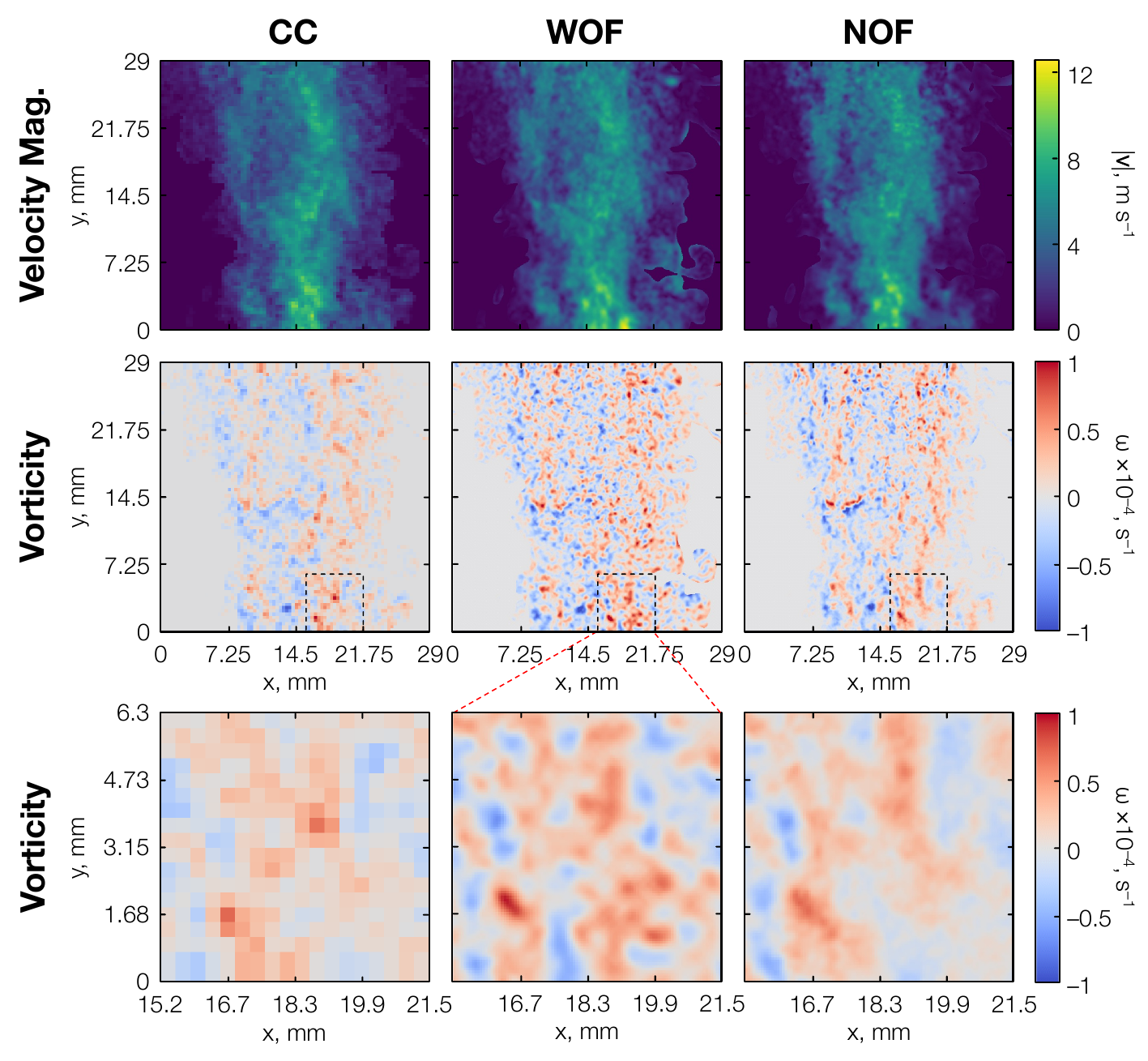}
    \caption{Reconstructed jet flow velocity magnitude fields (top), vorticity fields (middle), and magnified vorticity inserts (bottom).}
    \vspace*{-0.66em}
    \label{fig:jet panel}
\end{figure}

Figure~\ref{fig:jet panel} displays the velocity magnitude and vorticity fields, along with a close-up of the vorticity field within a region of interest, highlighted by the dashed box. To aid visualization, we masked the surrounding flow region outside the jet, where no particles were present (see Fig.~\ref{fig:cases}). These masks were generated by Gaussian blurring the particle images and thresholding them with an empirically determined value.\par

As shown in Fig.\ref{fig:jet panel}, WOF exhibits pronounced edge effects where the flow transitions from tracer-rich to tracer-free regions. A broader comparison of the vorticity fields reveals that NOF and CC produce more similar structures, while WOF introduces apparent spurious fluctuations, similar to those observed in Sec.~\ref{sec:results:3D HIT}. The close-up highlights an edge artifact unique to WOF but absent in both the NOF and CC results. Given the lack of particle information in this area (see Fig.~\ref{fig:cases}), it is reasonable to infer that NOF and CC provide more reliable estimates at the jet periphery. Artifacts in the WOF velocity field likely stem from sensitivity to variable seeding density and broader edge-related issues, as also seen in Sec.~\ref{sec:results:2D HIT}. While it is difficult to definitively identify the best method in experimental settings, WOF and NOF seem to exhibit broadly similar error characteristics: consistent with the synthetic results in Sec.~\ref{sec:results:3D HIT}, given the comparable level of turbulence. Further investigation is needed to quantify how measurement noise affects NOF performance relative to state-of-the-art methods like WOF, in order to clarify their respective strengths and limitations in real-world applications.\par

\section{Conclusions}
\label{sec:conclusions}
This manuscript introduces \emph{neural optical flow} (NOF), a neural-implicit method for optical flow (OF) particle image velocimery (PIV) aimed at improving velocity reconstruction for turbulent flows while providing a robust framework for data assimilation in PIV. NOF employs a coordinate-based neural network and a differentiable, nonlinear image-warping operator, eliminating the need for a multi-resolution scheme. Its neural-implicit representation simplifies data assimilation and enables direct pressure reconstruction for 2D and pseudo-2D flows. Furthermore, NOF offers an integrated stereo PIV framework that overcomes the limitations of traditional stereo PIV algorithms, which rely on stitching together separate 2D2C velocity field estimates.\par

We evaluated NOF on both synthetic and experimental datasets, comparing its performance to wavelet-based optical flow (WOF) and cross-correlation (CC), two state-of-the-art PIV techniques. In summary, our key findings are:
\begin{enumerate}[topsep=.5ex, itemsep=-.5ex, partopsep=.5ex, parsep=.5ex]
    \item Vanilla NOF provides slightly better velocity reconstructions compared to WOF in planar PIV due to its nonlinear image-warping scheme and inherent suppression of spurious high-frequency content.

    \item Applying a hard constraint on mass continuity for incompressible 2D\slash pseudo-2D flows significantly enhances reconstruction quality by constraining NOF to solenoidal solutions, reducing artifacts at high wavenumbers, particularly above the Nyquist wavenumber.
    
    \item NOF's all-in-one stereo reconstruction framework eliminates the error-amplifying recombination step required in traditional stereo algorithms. In our tests, this approach limited the increase in error when going from 2D2C to 2D3C estimates to just 0.8\%, compared to 4.4\% for WOF. Moreover, the out-of-plane energy spectra of NOF are free of spurious high-frequency artifacts, further validating the accuracy of our technique.
    
    \item NOF enables the direct inference of latent fields such as pressure, which is demonstrated in Sec.~\ref{sec:results:cylinder flow}. Direct pressure reconstruction can improve flow estimation through data assimilation, as quantified in Table~\ref{tab:2D HIT errors}, which also provides a robust alternative to post-processing with a pressure Poisson solver.
\end{enumerate}

In summary, NOF offers a versatile OF framework for PIV, addressing a broad range of fluid measurement challenges, including stereo PIV and pressure from PIV. While its performance matches or slightly surpasses existing state-of-the-art methods in simpler cases, NOF provides marked advantages beyond these performance gains for complex flows and stereo imaging. Additional benefits include smoother reconstructions in situations with limited, high-frequency data, enhanced feature extraction, and more accurate derivative-based flow statistics, especially in turbulent flows.\par

\appendix
\renewcommand{\thesection}{Appendix \Alph{section}}

\section{The Camera Transform}
\label{app:camera}

\subsection{Forward Transform}
\label{app:camera:forward}
Projecting points from world space to image space requires a camera model, with the pinhole camera often providing sufficient accuracy for PIV imaging. For a distortion-free pinhole model, the forward camera transform is defined as
\begin{equation}
    \label{equ:pinhole}
    \mathbf{s} = \zeta^{-1} \mathbf{K}
    \left(\mathbf{R}\mathbf{x} + \mathbf{t}\right).
\end{equation}
where $\mathbf{K}$ is the $2 \times 3$ intrinsic matrix, $\mathbf{R}$ is a $3 \times 3$ orthonormal rotation matrix, and $\mathbf{t}$ is the translation vector. This vector is defined as $\mathbf{t} = -\mathbf{Rc}$, where $\mathbf{c}$ is the camera's position in world coordinates. The scale factor $\zeta$ ensures projection onto the image plane and is defined as
\begin{equation}
    \zeta = \mathbf{R}_{3,*}\mathbf{x} + t_3,
\end{equation}
where $\mathbf{R}_{3,*}$ is the third row of $\mathbf{R}$ and $t_3$ is the third element of $\mathbf{t}$.\par

The intrinsic parameters in $\mathbf{K}$ include the distance from the pinhole to the sensor, $\hat{f}$, which defines the focal length of a pinhole camera, and the principal point, $\mathbf{s}_0 = (s_{0,1}, s_{0,2})$, where the optical axis intersects the sensor. Using these elements, the intrinsic matrix is
\begin{equation}
    \label{equ:intrinsic matrix}
    \mathbf{K} =
    \begin{bmatrix}
        \psi_1 \hat{f} & 0 & s_{0,1}\\
        0 & \psi_2 \hat{f} & s_{0,2}
    \end{bmatrix}.
\end{equation}
The terms $\psi_1 \hat{f}$ and $\psi_2 \hat{f}$ represent the effective focal lengths in pixel units along the horizontal and vertical directions; for square pixels, $\psi_1 = \psi_2 = \psi$. Note that the focal length of a pinhole camera, $\hat{f}$, differs from that of a physical lens, $f$. The relationship between these quantities is given by
\begin{equation}
    \label{equ:focal length relation}
    \hat{f} = f \left(1 + \frac{f}{d_\mathrm{o} - f}\right),
\end{equation}
where $d_\mathrm{o}$ is the distance from the aperture (or pinhole) to the object plane \cite{Wieneke2005}.\par

Lens distortions can be incorporated using a distortion model, typically applied to the normalized image coordinates, i.e., before scaling by the intrinsic matrix,
\begin{equation}
    \label{equ:normalized position}
    \mathbf{s}^\prime = \zeta^{-1}\left(\mathbf{Rx} + \mathbf{t}\right),
\end{equation}
where $(\cdot)^\prime$ indicates a normalized quantity and $\mathbf{s} = \mathbf{Ks}^\prime$. Radial and tangential distortions are computed using Brown’s polynomial distortion model \cite{Brown1996, Brown1971},
\begin{equation}
    \label{equ:distortion model}
    \mathbf{s}_\mathrm{dist}^\prime = \mathsf{D}\mathopen{}\left(\mathbf{s}^\prime\right).
\end{equation}
This function includes additional intrinsic parameters: specifically, a set of radial and tangential distortion coefficients. Bringing these elements together, the transform for a pinhole camera with distortions is
\begin{equation}
    \label{equ:forward transform}
    \mathbf{s} = \underbrace{\mathbf{K} \,\mathsf{D}\mathopen{}\left[ \zeta^{-1}\mathopen{}\left(\mathbf{R}\mathrm{x}+\mathbf{t}\right) \right]}_{\mathsf{\Psi}(\mathbf{x})}.
\end{equation}
In practice, the intrinsic and extrinsic parameters in $\mathbf{K}$, $\mathbf{R}$, $\mathbf{t}$, and $\mathsf{D}$ must be obtained through a camera calibration procedure, while $\psi$ is a fixed physical property of the sensor.\par

\subsection{Inverse Transform}
\label{app:camera:inverse}
The inverse mapping from sensor coordinates to world coordinates, $\mathbf{s}$ to $\mathbf{x}$, involves inverting the forward projection function in Eq.~\eqref{equ:forward transform}. Each pixel corresponds to a line through world space, and specifying the distance along this line, $l$, determines a unique 3D position. In planar and stereo PIV, this is the distance where the line intersects the laser sheet, as discussed in the following sections.\par

The general inverse transform is
\begin{equation}
    \label{equ:inverse transform}
    \mathbf{x} = \underbrace{\frac{l}{ \left\lVert \mathbf{r} \right\rVert_2} \overbrace{\mathbf{R}^\top \begin{bmatrix} \mathbf{K}^{-1}\mathbf{s} - \mathbf{s}_0^\prime \\ 1 \end{bmatrix}}^{\mathbf{r}} + \mathbf{c}}_{\mathsf{\Psi}^{-1}(\mathbf{s},l)},
\end{equation}
where $\mathbf{r}$ is a ray-direction vector; $\mathbf{K}^{-1}$ and $\mathbf{s}_0^\prime$ are defined as
\begin{equation}
    \mathbf{K}^{-1} \equiv
    \hat{f}^{-1} \begin{bmatrix}
        \psi_1^{-1} & 0 \\
        0 & \psi_2^{-1}
    \end{bmatrix}
    \quad\text{and}\quad
    \mathbf{s}_0^\prime \equiv \hat{f}^{-1}
    \begin{bmatrix}
        s_{0,1}/\psi_1 \\
        s_{0,2}/\psi_2
    \end{bmatrix}.
\end{equation}
When lens distortions are significant, the undistorted sensor coordinates must be found by inverting the distortion model,
\begin{equation}
    \label{equ:inverting distortion model}
    \mathbf{s}_\mathrm{dist} = \mathbf{K} \,\mathsf{D}\mathopen{}\left(\mathbf{K}^{-1}\mathbf{s} - \mathbf{s}_0^\prime\right),
\end{equation}
which is iteratively solved. Here, $\mathbf{s}_\mathrm{dist} = \mathbf{Ks}_\mathrm{dist}^\prime$ is a pixel centroid and the solution $\mathbf{s}$ is used in Eq.~\eqref{equ:inverse transform}.\par

\subsection{Application to Planar PIV}
\label{app:camera:planar}
In planar PIV, the camera is positioned perpendicular to the laser sheet, with the optical axis parallel to the $x_3$-axis of the world coordinate system. The world origin is set at the center of the laser sheet, and the camera's position is $\mathbf{c} = (0, 0, d_\mathrm{o})$, where $d_\mathrm{o}$ is the object distance. The camera is aimed towards the origin,
\begin{equation}
    \mathbf{R} =\begin{bmatrix}
        1&0&0 \\
        0&1&0 \\
        0&0&-1
    \end{bmatrix}
\end{equation}
\par

Assuming no distortions, and assuming that the measurement domain lies within the $x_1{-}x_2$ plane (i.e., $x_3 = 0$), the forward camera transform simplifies to
\begin{subequations}
    \label{equ:planar transform}
    \begin{align}
        \label{equ:planar transform:forward}
        \mathbf{s} &= \left(\psi_1 \hat{f} \frac{x_1}{d_\mathrm{o}} + s_{0,1}, \;\psi_2 \hat{f} \frac{x_2}{d_\mathrm{o}} + s_{0,2}\right)
        \intertext{and the inverse transform is given by}
        \label{equ:planar transform:inverse}
        \mathbf{x} &=  \frac{d_\mathrm{o}}{\hat{f}}\left(\frac{s_1-s_{0,1}}{\psi_1}, \;\frac{s_2-s_{0,2}}{\psi_2}, \;0\right).
    \end{align}
\end{subequations}
While minor misalignments and lens distortions are inevitable in a real experiment, they can often be neglected in planar PIV. The full transforms in \ref{app:camera:forward} and \ref{app:camera:inverse} can be used to handle cases with significant alignment and distortion issues.\par

\subsection{Application to Stereo PIV}
\label{app:camera:stereo}

\subsubsection{Translational Configuration}
\label{app:camera:stereo:translational}
The translational stereo PIV configuration builds on conventional planar PIV by using two cameras positioned to capture the laser sheet from slightly different perspectives. The object plane, lens plane, and sensor planes of both cameras are all parallel. In a symmetric configuration (see Fig.~\ref{fig:PIV}), the left and right camera positions in world space are
\begin{subequations}
    \label{equ:stereo camera position}
    \begin{align}
        \mathbf{c}_\mathrm{L} =& \left(-\frac{L}{2}, \;0, \;d_\mathrm{o}\right) \label{equ:stereo camera position:L}
        \intertext{and}
        \mathbf{c}_\mathrm{R} =& \left(\frac{L}{2}, \;0, \;d_\mathrm{o}\right), \label{equ:stereo camera position:R}
    \end{align}
\end{subequations}
where $L$ is the distance between the camera lenses, and subscripts ``L'' and ``R'' indicate the left and right cameras. Assuming perfect alignment with no lens distortion, $\mathbf{R} = \mathbf{I}$ and the forward camera transforms become
\begin{subequations}
    \label{equ:forward transform stereo trans}
    \begin{align}
            \mathbf{s}_\mathrm{L} =& \left(\psi_1 \hat{f} \frac{x_1 + L/2}{x_3 + d_\mathrm{o}} + s_{0,1}, \;\psi_2 \hat{f}\frac{x_2}{x_3 + d_\mathrm{o}} + s_{0,2}\right)
            \intertext{and}
            \mathbf{s}_\mathrm{R} =& \left(\psi_1 \hat{f} \frac{x_1 - L/2}{x_3 + d_\mathrm{o}} + s_{0,1}, \;\psi_2 \hat{f}\frac{x_2}{x_3 + d_\mathrm{o}} + s_{0,2}\right)
    \end{align}
\end{subequations}
These equations are consistent with previous derivations \cite{Arroyo1991, Raffel2018}. The inverse transforms, which map image coordinates back to world coordinates, are given by
\begin{subequations}
    \label{equ:inverse transform stereo trans}
        \begin{align}
            \mathbf{x}_\mathrm{L} =& \left[ \left(d_\mathrm{o} + x_3\right) \frac{s_1-s_{0,1}}{\psi_1 \hat{f}} - \frac{L}{2}, \;\left(d_\mathrm{o} + x_3\right)\frac{s_2-s_{0,2}}{\psi_2 \hat{f}}, \;x_3\right]
            \intertext{and}
            \mathbf{x}_\mathrm{R} =& \left[ \left(d_\mathrm{o} + x_3\right) \frac{s_1-s_{0,1}}{\psi_1 \hat{f}} + \frac{L}{2}, \;\left(d_\mathrm{o} + x_3\right)\frac{s_2-s_{0,2}}{\psi_2 \hat{f}}, \;x_3\right].
    \end{align}
\end{subequations}
While particles may be advected out of the measurement plane ($x_3 \neq 0$), stereo PIV assumes that the particles remain within the laser sheet, so we restrict all initial particle positions to the $x_1{-}x_2$ plane, with $x_3 = 0$. This simplification aligns with the planar assumption commonly used in stereo PIV.\par

Stereo PIV combines two 2D2C displacement fields, computed from each camera view using a method like CC or OF, into a single 2D3C displacement field in world space. Assuming square pixels, $\psi_1 = \psi_2 = \psi$, the relationship between the physical vector, $\Delta \mathbf{x} = (\Delta x_1, \Delta x_2, \Delta x_3)$, and the image displacement vectors, $\Delta \mathbf{s}_\mathrm{L}$ and $\Delta \mathbf{s}_\mathrm{R}$, can be expressed as
\begin{subequations}
\label{equ:stereo:displacement transform}
    \begin{align}
        \Delta x_1 =& \frac{\Delta s_{\mathrm{R},1} \left(x_1+L/2\right) - \Delta s_{\mathrm{L},1} \left(x_1-L/2\right)}{\psi ML + \left(\Delta s_{\mathrm{L},1} - \Delta s_{\mathrm{R},1}\right)} \label{equ:stereo:displacement transform:x1}\\
        \Delta x_2 =& \frac{x_2 \Delta x_3}{d_\mathrm{o}} + \frac{\Delta s_{\mathrm{L},2} + \Delta s_{\mathrm{R},2}}{2\psi M}\mathopen{}\left(\frac{\Delta x_3}{d_\mathrm{o}} + 1\right) \label{equ:stereo:displacement transform:x2}\\
        \Delta x_3 =& \frac{-d_\mathrm{o}\mathopen{}\left(\Delta s_{\mathrm{L},1} - \Delta s_{\mathrm{R},1}\right)}{\psi ML + \left(\Delta s_{\mathrm{L},1} - \Delta s_{\mathrm{R},1}\right)}. \label{equ:stereo:displacement transform:x3}
    \end{align}
\end{subequations}
Since each 3C displacement vector in world space corresponds to two 2C displacement vectors in image space (four pieces of information), the stereo analysis is overdetermined. This redundancy can be leveraged to improve the accuracy of Eq.~\eqref{equ:stereo:displacement transform:x2}, e.g., by averaging $\Delta s_{\mathrm{L},2}$ and $\Delta s_{\mathrm{R},2}$. This is the origin of the asymmetry between Eqs.~\eqref{equ:stereo:displacement transform:x1} and \eqref{equ:stereo:displacement transform:x2}.\par

\subsubsection{Rotational Configuration}
\label{app:camera:stereo:angular}
Rotational stereo PIV employs two inward-facing cameras that are aimed at a point in the laser sheet. Scheimpflug adapters are commonly used to ensure that the entire sheet remains in focus.\footnote{Rotational configurations can be implemented without Scheimpflug adapters, although the aperture must be stopped down as far as possible to achieve a sufficient depth-of-focus \cite{Westerweel1991, Prasad1993, Lawson1997}.} When the Scheimpflug condition is met, the object, lens, and sensor planes all intersect along a common line. This contrasts with the standard pinhole model in Eq.~\eqref{equ:pinhole}, which assumes parallel alignment between the lens and sensor planes.\par

Cornic et al. \cite{Cornic2016} developed a modified pinhole model to account for the sensor tilt introduced by a Scheimpflug adapter,
\begin{equation}
    \label{equ:pinhole scheimflug}
    \mathbf{s} = \zeta^{-1} \mathbf{K} \mathbf{S} \mathbf{R}_\alpha \left(\mathbf{Rx} + \mathbf{t}\right),
\end{equation}
where $\mathbf{R}_\alpha$ is the rotation matrix accounting for sensor tilt, $\mathbf{S}$ is the Scheimpflug matrix, and
\begin{equation}
    \zeta = (0,0,1) \cdot \left[\mathbf{S}\mathbf{R}_\alpha \left(\mathbf{Rx} + \mathbf{t}\right)\right]
\end{equation}
is the scale factor that projects world coordinates onto the image plane. The sensor rotation matrix, $\mathbf{R}_\alpha$, can be decomposed into rotations about the $s_1$ and $s_2$ axes, with respective tilt angles $\alpha_1$ and $\alpha_2$,
\begin{equation}
    \label{equ:rotation sensor}
    \mathbf{R}_\alpha = \begin{bmatrix}
        \cos(\alpha_2) & 0 & \sin(\alpha_2)\\
        0 & 1 & 0\\
        -\sin(\alpha_2) & 0 & \cos(\alpha_2)
    \end{bmatrix} \begin{bmatrix}
        1 & 0 & 0\\
        0 & \cos(\alpha_1) & \sin(\alpha_1)\\
        0 & -\sin(\alpha_1) & \cos(\alpha_1)       
    \end{bmatrix}.
\end{equation}
The Scheimpflug matrix, $\mathbf{S}$, depends on these same tilt angles,
\begin{equation}
    \label{equ:Scheimflug array}
    \mathbf{S} = \begin{bmatrix}
        \cos(\alpha_1)\cos(\alpha_2) & 0 & \sin(\alpha_2)\\
        0 & \cos(\alpha_1)\cos(\alpha_2) & -\cos(\alpha_2)\sin(\alpha_1)\\
        0 & 0 & 1\\
    \end{bmatrix}.
\end{equation}
Lens distortions are modeled similarly to Eq.~\eqref{equ:distortion model}. Incorporating the distortion model, the forward camera transform becomes
\begin{equation}
    \label{equ:forward transform Scheimflug}
    \mathbf{s} = \mathbf{K} \,\mathsf{D}\mathopen{}\left[\zeta^{-1} \mathbf{S}\mathbf{R}_\alpha \left(\mathbf{Rx} + \mathbf{t}\right)\right].
\end{equation}\par

For the inverse transform, where the goal is to map from sensor coordinates to world coordinates, we assume a laser sheet at the origin with a normal vector of $(0, 0, 1)$. The resultant inverse mapping is
\begin{subequations}
    \label{equ:inverse transform Scheimflug}
    \begin{align}
        \mathbf{x} &= -\frac{c_{3}}{r_3} \mathbf{r} + \mathbf{c},
        \intertext{where the ray direction is}
        \mathbf{r} &= \mathbf{R}^\top \mathbf{R}_\alpha^\top \mathbf{S}^{-1} \begin{bmatrix} \mathbf{K}^{-1}\mathbf{s} - \mathbf{s}_0^\prime \\ 1 \end{bmatrix},
    \end{align}
\end{subequations}
Here, $c_3$ and $r_3$ are the third components of $\mathbf{c}$ and $\mathbf{r}$, respectively. As before, if lens distortions are present, the distortion model must be inverted to recover the undistorted sensor coordinates using Eq.~\eqref{equ:inverting distortion model}.\par

\section{Network Architecture}
\label{app:arch}
In this work, we use a coordinate neural network to represent flow fields in $\mathcal{P}$. The network consists of an input layer, an output layer, and a series of $n_\mathrm{l}$ hidden layers,
\begin{subequations}
    \label{equ:method:architecture}
    \begin{align}
        \mathbf{z}_\mathrm{out} &= \overbrace{\mathbf{W}^{n_\mathrm{l}+1}\mathopen{}\left[\mathsf{L}^{n_\mathrm{l}} \circ \mathsf{L}^{n_\mathrm{l}-1} \circ \dots \circ \mathsf{L}^2 \circ \mathsf{F}\mathopen{}\left(\mathbf{z}_\mathrm{in}\right)\right]+\mathbf{b}^{n_\mathrm{l}+1}}^{\mathsf{N}\mathopen{}\left(\mathbf{z}_\mathrm{in}\right)},
        \intertext{with}
        \mathbf{z}^l &= \underbrace{\mathrm{GELU}\mathopen{}\left(\mathbf{W}^l\mathbf{z}^{l-1} + \mathbf{b}^l\right)}_{\mathsf{L}^l\mathopen{}\left(\mathbf{z}^{l-1}\right)} \quad\text{for}\quad l \in \{2, 3, \dots, n_\mathrm{l}\}.
    \end{align}
\end{subequations}
In this paper, $\mathbf{z}_\mathrm{in}$ is $(\mathbf{x},t)$ and $\mathbf{z}_\mathrm{out}$ is $\mathbf{v}$.\footnote{The output $\mathbf{z}_\mathrm{out}$ is $\varphi$ or $\boldsymbol\upvarphi$ for 2D and 3D velocity fields with a hard constraint on continuity, as discussed in Sec.~\ref{sec:method:physics-based}, and pressure is outputted when using a Navier--Stokes penalty term, $\mathbf{z}_\mathrm{out} = (\mathbf{v}, p)$, $(\varphi, p)$, or $(\boldsymbol\upvarphi, p)$.} The vector $\mathbf{z}^l$ contains the values of neurons in the $l$th layer, $\mathbf{W}^l$ and $\mathbf{b}^l$ are the weight matrix and bias vector, and $\mathrm{GELU}$, which stands for 
Gaussian error linear unit, is a nonlinear activation function that is applied element-wise. We also use the weight normalization technique of Salimans and Kingma \cite{Salimans2016} to accelerate training.\par

To mitigate the low-frequency spectral bias inherent in gradient-based optimizers \cite{Wang2021}, we replace the first hidden layer, $\mathsf{L}^1$, with a Fourier encoding layer \cite{Tancik2020},
\begin{equation}
    \label{equ:method:FE}
    \mathbf{z}^1 = \underbrace{\left[\sin\mathopen{}\left(2\pi \mathbf{f}_1 \cdot \mathbf{z}^0\right), \,\cos\mathopen{}\left(2\pi \mathbf{f}_1 \cdot \mathbf{z}^0\right), \dots, \,\sin\mathopen{}\left(2\pi \mathbf{f}_{n_\mathrm{f}} \cdot \mathbf{z}^0\right), \,\cos\mathopen{}\left(2\pi \mathbf{f}_{n_\mathrm{f}} \cdot \mathbf{z}^0\right)\right]}_{\mathsf{F}\mathopen{}\left(\mathbf{z}_\mathrm{in}\right)}
\end{equation}
Here, $n_\mathrm{f}$ is the number of Fourier features and $\mathbf{f} = (f_1, f_2, f_3)$ are vectors of random frequencies, sampled from a normal distribution and fixed during training. As discussed in Sec.~\ref{sec:method:NOF}, the standard deviation of the frequency distribution is a source of implicit regularization, which can introduce a controlled spectral bias. For time-resolved data, the spatial and temporal standard deviations, i.e., for $(f_1,f_2)$ and $f_3$, respectively, are adjusted independently to account for variations in spectral content.\par

The network size (depth and width) is crucial for ensuring its ability to represent the target function. We tested architectures with 7--12 hidden layers and 200--600 neurons per layer, observing consistent results across all configurations. A network with nine hidden layers and 500 neurons per layer provided sufficient representation across all tested flow cases, and therefore was used throughout the study.

The Adam optimizer \cite{Zhang2018} was employed with a constant learning rate of $10^{-3}$. Training durations varied depending on the flow scenario and are reported throughout the paper on a per-case basis. All experiments were implemented using PyTorch~2.4 on an NVIDIA RTX 3090 GPU. Training times ranged anywhere from 3 minutes to 25 minutes depending on the variant used and complexity of the underlying field. The mean and standard deviation in execution times across 10 runs on the 2D HIT case for representative algorithms is shown in Table~\ref{tab:time}. Execution times follow similar patterns for other flow scenarios, however forming a robust method to estimate NOF execution times is difficult; unlike CC and WOF, whose time complexity scales with the size of the discrete image grid, NOF scales with some concept of information magnitude and density in the underlying velocity field.\par

\begin{table}[ht]
    \caption{Per-Frame Execution Time (in Seconds) for 2D HIT Reconstructions}
    \centering
    \tabcolsep=0.2cm
    \vspace*{.1em}
    \begin{tabular}{c c c c c}
        \hline\hline \\[-.75em]
        \multirow{2}{*}{} & \multicolumn{4}{c}{\textbf{Algorithm}}\\
        & \multicolumn{1}{c}{CC} & \multicolumn{1}{c}{WOF} & \multicolumn{1}{c}{NOF} & \multicolumn{1}{c}{NOF-HD}\\ 
        \hline 
        \multirow{1}{*}{}  &  &  &  &\\[-.75em]
        \multirow{1}{*}{} Mean time & $\mathbf{0.32}$ & $3.71$ & $233.53$ & $277.63$\\
        \multirow{1}{*}{} Std. dev. & $\mathbf{4.5\times10^{-3}}$ & $2.9\times10^{-2}$ & $21.39$ & $7.92$\\
        \hline\hline
    \end{tabular}
    \label{tab:time}
\end{table}

\section{Interpolation Schemes}
\label{app:interp}
Approximating the integral in Eq.~\eqref{equ:NOF:data loss:int} requires continuous sampling of the intensity field, which is inherently discrete in image data. To address this, we tested both bilinear and bicubic interpolation schemes, each offering differentiable representations of the intensity field. These interpolants are essential for ensuring the accuracy and stability of our image warping operator during optimization.\par

Let $\mathbf{s}_{i,j}$ be the centroid of the $(i,j)$th pixel and $I_{i,j}$ be its intensity. The image vector for an $m \times n$ image is $\mathbf{I} = \{I_{i,j}\mid i = 1, \dots, m, j = 1, \dots, n\}$. Assuming pixels with a side length of one, the bilinear interpolation operator $\mathsf{B}$ is
\begin{subequations}
    \begin{align}
        \hat{I}\mathopen{}\left(\mathbf{s}\right) &= \overbrace{w_1 I_{i,j} + w_2 I_{i+1,j} + w_3 I_{i,j+1} + w_4 I_{i+1,j+1}}^{\mathsf{B}\mathopen{}\left(\mathbf{I}, \mathbf{s}\right)},
        \intertext{where the interpolation weights are}
        w_1 &= \left(1 - s_1 + s_{i,j,1}\right) \left(1 - s_2 + s_{i,j,2}\right) \\
        w_2 &= \left(s_1 - s_{i,j,1}\right) \left(1 - s_2 + s_{i,j,2}\right) \\
        w_3 &= \left(1 - s_1 + s_{i,j,1}\right) \left(s_2 - s_{i,j,2}\right) \\
        w_4 &= \left(s_1 - s_{i,j,1}\right) \left(s_2 - s_{i,j,2}\right),
    \end{align}
\end{subequations}
and where $\mathbf{s}$ is the query point, enclosed by the centroids $\mathbf{s}_{i,j}$, $\mathbf{s}_{i+1,j}$, $\mathbf{s}_{i,j+1}$, and $\mathbf{s}_{i+1,j+1}$. This method provides a suitable approximation of intensity across the image.\par

Bicubic interpolation, on the other hand, uses a 16-pixel neighborhood and computes weights based on polynomial functions of the distances between the query point and pixel centroids. This method captures more detail and provides a higher-order representation than bilinear interpolation, which can be especially important near the edges of an image. We use the convolution-based implementation from Keys \cite{Keys1981}, setting the interpolation parameter to $-0.75$, as recommended for optimal smoothness. Further technical details on both interpolation schemes can be found in \emph{Numerical Recipes} \cite{Press2007}.\par

In our implementation, we use the \texttt{grid\_sample} function in PyTorch for both forms of interpolation. Preliminary tests indicated that bilinear interpolation offers sufficient accuracy at lower computational cost, so it is used as the default throughout this work, unless otherwise stated.\par

\section*{Acknowledgements}
This research was supported by Mitsubishi Electric Research Laboratories. S.J.G. acknowledges support from FAU Erlangen-N{\"u}rnberg. J.P.M. acknowledges support from the DoD through an NDSEG fellowship.\par


\end{document}